\documentclass[aps,pra,twocolumn,superscriptaddress,a4paper,noeprint]{revtex4-2}
\usepackage{graphicx}

\usepackage{graphicx}
\usepackage{amsmath}
\usepackage{float}
\usepackage{color}
\usepackage{upgreek}
\usepackage{graphicx, nicefrac}
\pdfoutput=1 %Needed for ArXiv submission

\usepackage[colorlinks=true,citecolor=blue,linkcolor=magenta]{hyperref}
\hypersetup{
pdfauthor = {},
colorlinks = true, linkcolor = blue, urlcolor=blue, bookmarksnumbered =  true}

\newcommand{\ket}[1]{\left|#1\right>}

\begin{document}
%Title of paper
\title{Widely-tunable, doubly-resonant Raman scattering on diamond in an open microcavity}
\author{Sigurd Fl\aa gan}
\affiliation{Department of Physics, University of Basel, Klingelbergstrasse 82, Basel CH-4056, Switzerland}
\author{Patrick Maletinsky}
\affiliation{Department of Physics, University of Basel, Klingelbergstrasse 82, Basel CH-4056, Switzerland}
\author{Richard J. Warburton}
\affiliation{Department of Physics, University of Basel, Klingelbergstrasse 82, Basel CH-4056, Switzerland}
\author{Daniel Riedel}
\affiliation{Department of Physics, University of Basel, Klingelbergstrasse 82, Basel CH-4056, Switzerland}
\affiliation{Currently at E. L. Ginzton Laboratory, Stanford University,
Stanford, CA 94305, USA}
\email{Corresponding author: riedeld@stanford.edu}
\date{\today}
\begin{abstract}
Raman lasers based on bulk diamond are a valuable resource for generating coherent light in wavelength regimes where no common laser diodes are available. Nevertheless, the widespread use of such lasers is limited by their high threshold power requirements on the order of several Watts. Using on-chip microresonators, a significant reduction of the lasing threshold by more than two orders of magnitude has been shown. However, these resonators lack a continuous tuning mechanism and, mainly due to fabrication limitations, their implementation in the visible remains elusive. Here, we propose a platform for a diamond Raman laser in the visible. The device is based on a miniaturized, open-access Fabry-Perot cavity. Our microcavity provides widely-tunable doubly-resonant enhancement of Raman scattering from high quality single-crystalline diamond. 
We demonstrate a $>$THz continuous tuning range of doubly-resonant Raman scattering, a range limited only by the reflective stopband of the mirrors. Based on the experimentally determined quality factors exceeding 300\,000, our theoretical analysis suggests that, with realistic improvements, a sub-mW threshold is readily within reach. Our findings pave the way to the creation of a universal low-power frequency shifter, a potentially valuable addition to the nonlinear optics toolbox.
\end{abstract}

\maketitle
%%%%%%%%%%%%%%%%%%%%%%%%%%%%%%%%%%%%%%%%%%%%%%%%%%%%%%%%%%%%%%%%%%%%%%%%%%%
% INTRODUCTION
%%%%%%%%%%%%%%%%%%%%%%%%%%%%%%%%%%%%%%%%%%%%%%%%%%%%%%%%%%%%%%%%%%%%%%%%%%%
\section{Introduction}
The development of new laser architectures to create coherent radiation at arbitrary wavelengths is a continuous endeavor. Solid-state based lasers have emerged as the leading platform due to their compact design, reliability, highly stable output and beam quality\,\cite{Svelto2010}. Several wavelength regimes, however, are not directly accessible with semiconductor lasers and require wavelength conversion in a nonlinear medium.

A promising technique relies on Raman scattering where, upon the creation of a phonon, photons are red-shifted by the fixed phonon energy. This energy shift and the efficiency of the process, quantified by the Raman gain parameter, depend on the material and in particular on the phonons involved\,\cite{Raman1928}. The main advantage of this approach is that in principle any laser wavelength can be achieved if a suitable pump laser is available\,\cite{Pask2003}. This gain mechanism differs from stimulated emission from excited dopants which exhibit a fixed gain frequency bandwidth\,\cite{Williams2018}.

Diamond is particularly well suited to the creation of a Raman-based laser due to its exceptional properties\,\cite{Mildren2013}, notably a large Raman gain ($\sim75\,\textrm{GW}\cdot\textrm{cm}^{-1}$ at 532\,nm)\,\cite{Mildren2013} and a large Raman shift ($\sim$1\,332\,cm$^{-1}$)\,\cite{Zaitsev2010}. This large Raman shift enables wavelengths to be accessed for which no ideal solution exists in terms of cost, convenience and output power. Other advantages of diamond in this context are the wide bandgap of diamond, which prevents free carrier absorption minimizing optical losses in the visible and the ultraviolet wavelength regimes, and diamond's high thermal conductivity, which facilitates efficient heat management.

Diamond Raman lasers have enabled the creation of coherent radiation in exotic wavelength regimes, e.g.\ the yellow band\,\cite{Greentree2010, Spence2010,Yang2020,Mildren2008,Mildren2009,Sheng2019}, where no common laser diodes are available. High-power diamond Raman lasers have been implemented across a large range of wavelengths, from the ultraviolet\,\cite{Granados2011} across the visible\,\cite{Chrysalidis2019,Yang2020,Sheng2019} and infrared\,\cite{Lubeigt2010,Parrotta2013,Kitzler2012,Sabella2010,Sabella2011,Feve2011,Williams2014,Antipov2019, Casula2017}, all the way to the mid-infrared\,\cite{Sabella2010}. However, current implementations are limited by their high threshold pump power requirements, typically several Watts.

Micro- and nanophotonic engineering offers the potential of reducing the threshold for lasing. Through resonant recirculation of the pump beam in a cavity with a small mode cross-section, the intensity-dependent Raman gain is significantly enhanced. In addition, simultaneous coupling of the Raman field to a second cavity mode boosts the efficiency of stimulated emission. Based on this doubly-resonant configuration, low-threshold Raman lasers in the infrared wavelength regime were demonstrated in silica microspheres\,\cite{Spillane2002}. Chip-integration has been implemented using silica microtoroids\,\cite{Kippenberg2004}, silicon waveguides\,\cite{Rong2005,Rong2005b} and racetrack resonators\,\cite{Rong2008}. Combining high quality ($Q$) factors with small mode volumes ($V$), silicon photonic crystals have enabled ultra-low threshold Raman lasing in the nanowatts regime\,\cite{Takahashi2013}. Recently, molecules adsorbed to silica microtoroids emerged as promising gain medium\,\cite{Shen2020}. Advances in diamond nanofabrication enabled the demonstration of integrated diamond Raman lasers using ring resonators at infrared\,\cite{Latawiec2015} and near-visible\,\cite{Latawiec2018} wavelengths. Largely on account of material and nanofabrication constraints, Raman microlasers at visible frequencies, however, remain elusive.

%%%%%%%%%%%%%%%%%%%%%%%%%%%%%%%%%%%%%%%%%%%%%%%%%%%%%%%%%%%%%%%%%%%%%%%%%%%
% FIG. 1
%%%%%%%%%%%%%%%%%%%%%%%%%%%%%%%%%%%%%%%%%%%%%%%%%%%%%%%%%%%%%%%%%%%%%%%%%%%
\begin{figure}[t!]
\includegraphics[width=\columnwidth]{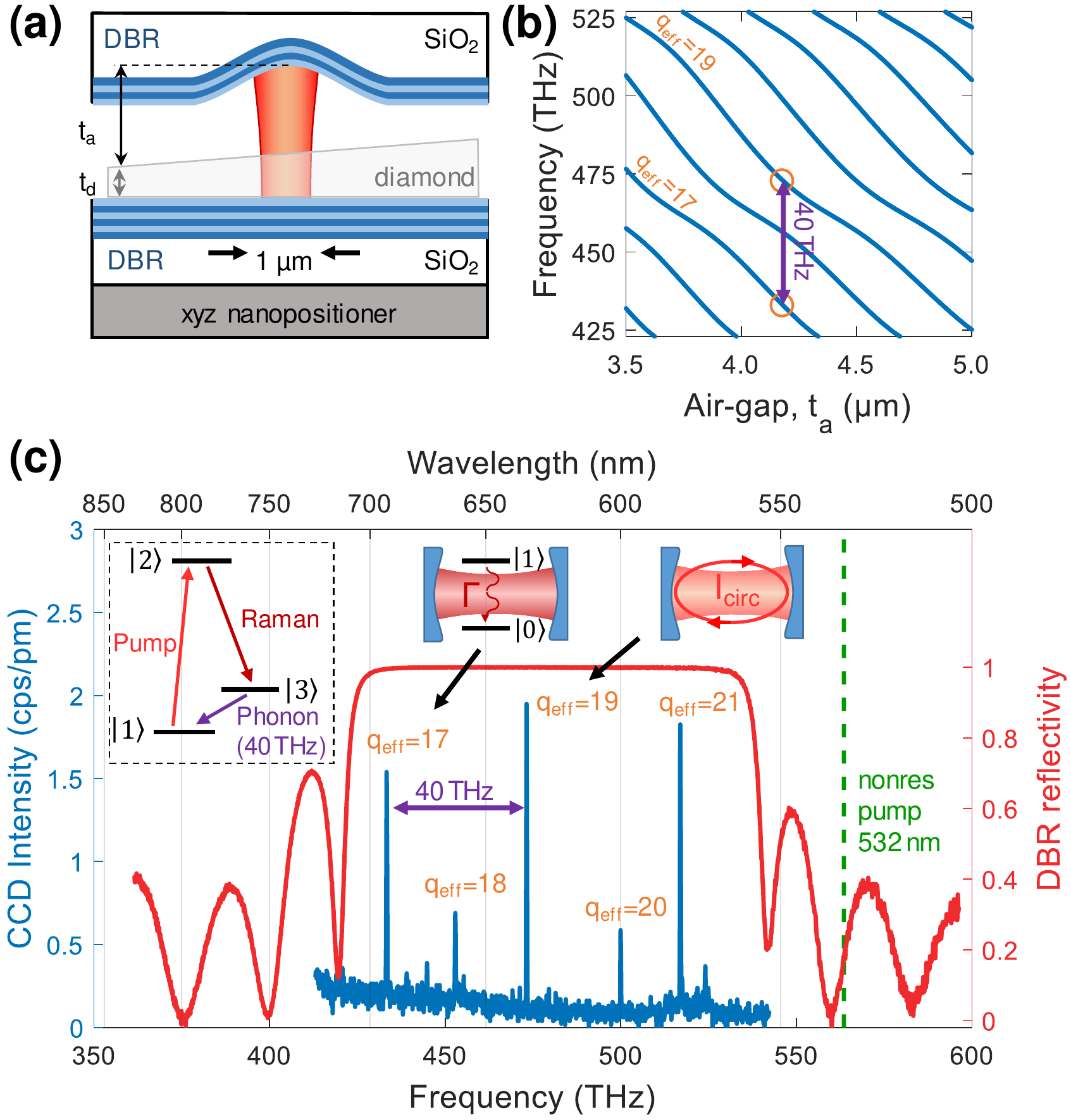}
\caption{\textbf{(a)} Schematic of the plano-concave Fabry-Perot microcavity. The cavity is formed by two fused silica (SiO$_2$) substrates coated with a distributed Bragg reflector (DBR). One of the substrates exhibits spherical microindentations resulting in a Gaussian resonator mode with a small beam waist ($\sim 1\,\upmu\textrm{m}$). Piezoelectric nanopositioners allow for spatial and spectral tunability of the cavity mode. The wedged diamond membrane enables the diamond thickness within the cavity mode to be changed by lateral positioning. The thickness gradient of the diamond is exaggerated for enhanced visibility.
\textbf{(b)} One-dimensional transfer-matrix calculation of the cavity mode-structure for a diamond thickness of $t_\textrm{d}=756\,\textrm{nm}$. The presence of the diamond membrane leads to a nonlinear mode dispersion. At a mirror separation of $\sim 4.18\,\upmu\textrm{m}$, the frequency spacing between modes $q_\textrm{eff}=17$ and $q_\textrm{eff}=19$ equals the frequency of the optical phonon in diamond ($\sim40\,$THz).
\textbf{(c)} In red: measurement of the reflectivity of the planar DBR revealing a stopband centered at 625\,nm with a bandwidth of $\sim100$\,THz. In blue: experimental cavity spectrum for a fixed cavity length under $532\,\textrm{nm}$ illumination. The diamond thickness and mirror separation can be inferred by comparing the spectrum with the simulated mode structure in (b). Inset: Raman process, depicted as a three-level system. A photon is converted to a red-shifted photon and an optical phonon of fixed frequency.}
\label{fig:Fig_1}
\end{figure}
%%%%%%%%%%%%%%%%%%%%%%%%%%%%%%%%%%%%%%%%%%%%%%%%%%%%%%%%%%%%%%%%%%%%%%%%%%%

We propose to resolve this conundrum by using a different platform for the implementation of a low-threshold diamond Raman laser, namely a highly miniaturized Fabry-Perot cavity\,\cite{Riedel2020}. Such microcavities, sometimes referred to as open microcavities, offer high $Q$/$V$ ratios and thereby promote strong light-matter interactions\,\cite{Najer2019,Wang2019}. We recently demonstrated that high $Q$-factors at visible wavelengths can be achieved when embedding a high-quality diamond membrane into the cavity\,\cite{Flagan2021}. In comparison to ring resonators, there are no limitations due to bending losses and surface roughness is less detrimental.

%%%%%%%%%%%%%%%%%%%%%%%%%%%%%%%%%%%%%%%%%%%%%%%%%%%%%%%%%%%%%%%%%%%%%%%%%%%
% Methods
%%%%%%%%%%%%%%%%%%%%%%%%%%%%%%%%%%%%%%%%%%%%%%%%%%%%%%%%%%%%%%%%%%%%%%%%%%%
\section{Methods}
Our plano-concave microcavity design supports Gaussian fundamental modes (Fig.\,\ref{fig:Fig_1}\,(a))\,\cite{Hunger2010, Muller2010, Barbour2011, Greuter2014,Janitz2015,Hummer2016}. The microcavity is formed by two mirror-coated fused silica substrates, one of which contains an array of spherical micromirrors fabricated via laser ablation\,\cite{Hunger2012} enabling efficient coupling to a single free-space mode\,\cite{Riedel2020, Tomm2021}. The radii of curvature of these micromirrors are $\sim10\,\upmu\textrm{m}$ resulting in a beam waist of $\sim1\,\upmu\textrm{m}$. We integrate high-quality single-crystalline diamond micromembranes ($\sim20\times20\times0.8\,\upmu\textrm{m}^3$) into the cavity using a micromanipulator\,\cite{Riedel2014,Riedel2017} (see Appendix\,\ref{sec:methods}).

For conventional Fabry-Perot resonators, the resonance wavelength changes linearly with the mirror separation $t_{\textrm{a}}$. However, the presence of a diamond membrane with thickness $t_{\textrm{d}}$ significantly alters this linear mode structure. The hybridization of modes confined in the air and diamond layers of the resonator manifests in avoided crossings (Fig.\,\ref{fig:Fig_1}\,(b))\,\cite{Riedel2017,Janitz2015,Haussler2019,Bogdanovic2017,Jensen2020,Ruf2021}. The cavity resonance frequencies depend on both the separation of the two mirrors and the thickness of the diamond at the location of the cavity mode (see Appendix\,\ref{sec:sim}).

One of the main advantages of our cavity platform is the \textit{in situ} tuning capability. Both the separation of the mirrors and the lateral position of the cavity mode with respect to the diamond membrane can be controlled via a stack of piezoelectric nano-positioners (attocube {ANPx51, ANPz51}). By including a slight thickness gradient into the diamond membrane, we are able to tune the exact diamond thickness of the membrane in the cavity by adjusting the relative lateral position of the mirrors.
The \textit{in situ} tuning capability allows us to control both the absolute frequency as well as the relative splitting of the resonator modes.

%Strong enhancement of Raman scattering requires the establishment of a double-resonance condition, i.e.\ a simultaneous resonance of one cavity mode with the pump photons, and resonance of another cavity mode with the Raman photons. 
Excitation of the first order Stokes process in diamond can be modelled as a three-level atom-like system (inset Fig.\,\ref{fig:Fig_1}\,(c)) involving a ground state $\ket{1}$, a virtual state $\ket{2}$ and a meta-stable state $\ket{3}$. A pump laser excites the ground-state population from $\ket{1}$ to $\ket{2}$. The system decays via state $\ket{3}$ emitting a red-shifted photon ($\ket{2}\rightarrow\ket{3}$) followed by an optical phonon of fixed frequency ($\sim40\,$THz, $\ket{3}\rightarrow\ket{1}$).

By coupling both the pump and the Stokes photons to a cavity mode, the Stokes process can be strongly enhanced\,\cite{Petrak2014}. It should be noted that no population inversion is required for stimulated Raman scattering and hence the creation of a Raman laser\,\cite{Pavlov2018}.
Importantly, the gain of the Raman process is maximized by strong confinement of and coupling between the pump and Raman modes, which suggests the use of fundamental resonator modes. Careful tuning of the mirror separation $t_\textrm{a}$ and the diamond thickness $t_\textrm{d}$ allows the double-resonance condition to be established for a wide range of pump wavelengths in the visible wavelength regime. When changing the pump wavelength $t_\textrm{d}$ and $t_\textrm{a}$ need to be adjusted such that the cavity both remains resonant with the pump laser while another mode is red-detuned exactly by the Raman shift (see Appendix\,\ref{sec:drestuning}).

The frequency of the Raman output is determined by the frequency of the pump laser and the fixed Raman shift. We use a continuous-wave (cw) narrow-band tunable red diode laser (Toptica DL Pro 635, $\lambda=630\dots640\,\textrm{nm}$) as a pump source. The operation wavelength range of the cavity is given by the reflective stopband of the DBR, which we determine using a white-light transmission measurement\,\cite{Riedel2020}. Figure\,\ref{fig:Fig_1}\,(c) displays the stopband of the planar bottom mirror which is centered around $\lambda_{\textrm{c,bot}}=625\,\textrm{nm}$; the reflectivity is more than 99\% over a bandwidth of $\sim 100$\,THz. The top mirror has similar properties but with a stopband centered at $\lambda_{\textrm{c,top}}=629\,\textrm{nm}$.

%%%%%%%%%%%%%%%%%%%%%%%%%%%%%%%%%%%%%%%%%%%%%%%%%%%%%%%%%%%%%%%%%%%%%%%%%%%
% RESULTS
%%%%%%%%%%%%%%%%%%%%%%%%%%%%%%%%%%%%%%%%%%%%%%%%%%%%%%%%%%%%%%%%%%%%%%%%%%%
\section{Results}

To characterize the mode structure of the cavity, we couple a cw green laser at $532\,\textrm{nm}$ into the cavity through the curved top mirror. We tune the mirror separation by applying a voltage to the piezo using a highly-stable voltage source (Basel Precision Instruments SP 927). Background photoluminescence (PL) from the diamond acts as an internal light source and couples to the different resonator modes\,\cite{Riedel2020}. Figure\,\ref{fig:Fig_1}\,(c) displays a PL spectrum collected through the top mirror. We set the mirror separation such that the splitting between the modes with effective mode numbers $q_\textrm{eff}=17$ and $q_\textrm{eff}=19$ corresponds to the Raman shift in diamond ($\Delta\nu_\textrm{R}\sim 40\,\textrm{THz}$). We define the effective mode number by the number of half wavelengths between the two mirrors, i.e.\ within the air-gap and the diamond layer, $q_\textrm{eff}\lambda/2\approx t_\textrm{a}+n_\textrm{d}t_\textrm{d}$. A small deviation from integer values of $q_\textrm{eff}$ is caused by field penetration into the DBR mirrors\,\cite{Kelkar2015,Koks2021}. Using a one-dimensional transfer-matrix calculation we infer the mirror separation and diamond layer thickness to be $t_\textrm{a}=4.18\,\upmu\textrm{m}$ and $t_\textrm{d}=756\,\textrm{nm}$, respectively (Figure\,\ref{fig:Fig_1}\,(b), Appendix\,\ref{sec:sim}).

\begin{figure}[t!]
\includegraphics[width=\columnwidth]{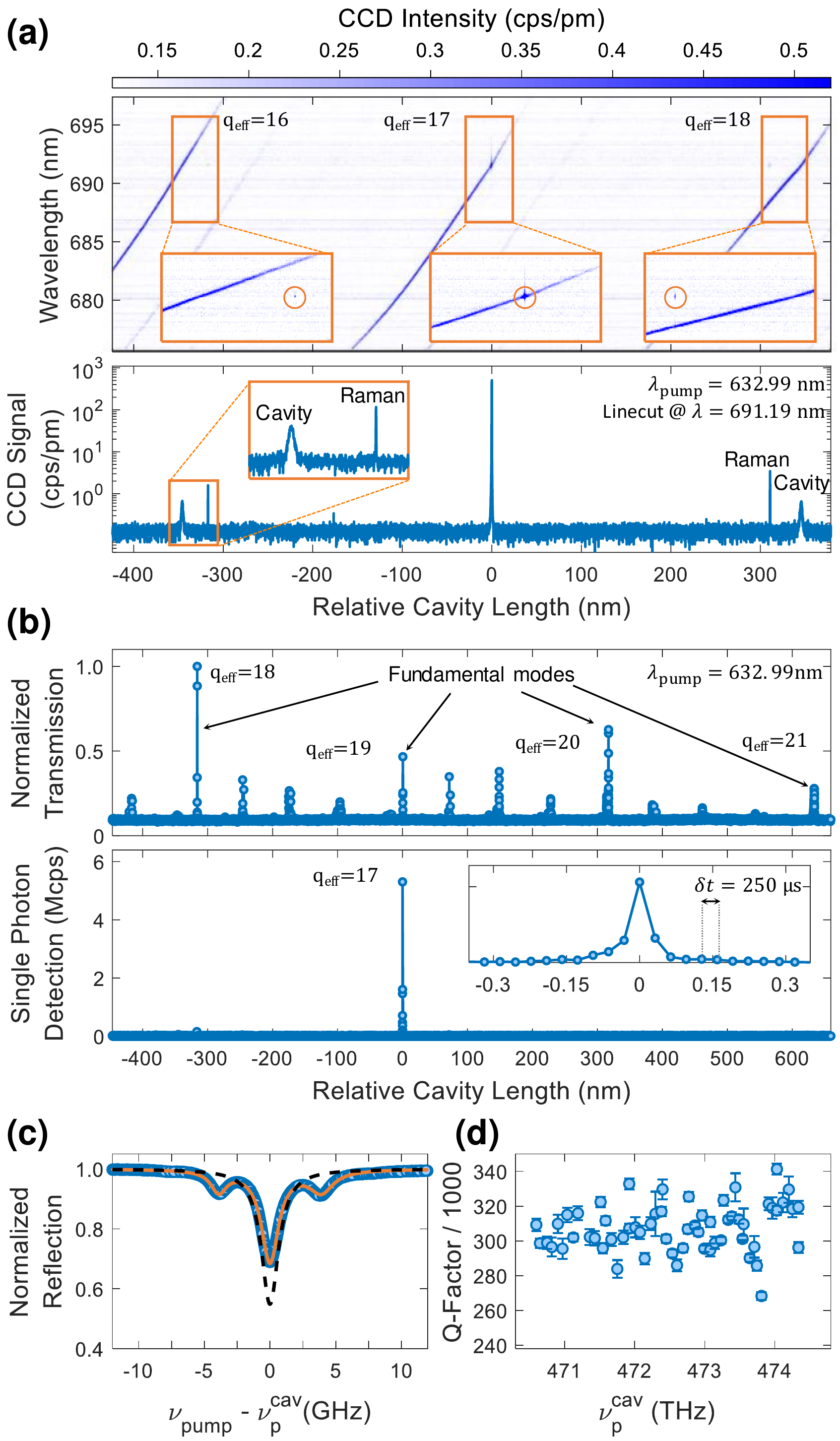}
\caption{Demonstration of cavity-enhanced doubly-resonant Raman scattering. \textbf{(a)} (top) Optical spectra as a function of relative cavity length reveal the cavity mode structure. Here, a laser at $532\,\textrm{nm}$ excites background PL in the diamond and a narrowband diode laser at $\lambda_\textrm{pump}=632.99\,\textrm{nm}$ drives Raman scattering. For $q_\textrm{eff}=19$ and $q_\textrm{eff}=17$, the pump and Raman scattered light are resonant simultaneously, i.e.\ the double resonance condition is met. (bottom) Linecut at $\lambda=691.19\,\textrm{nm}$ highlighting the strong signal enhancement of doubly-resonant Raman scattering.
\textbf{(b)} Cavity transmission (top) and cavity emission at wavelengths $>644\textrm{nm}$ measured with a single photon counter (bottom) as a function of cavity length for $\lambda_\textrm{pump}=632.99\,\textrm{nm}$. A cavity signal is only observed when the double resonance condition is satisfied. The piezo voltage is changed at a rate of $2\,\textrm{mV}/\textrm{ms}$.
\textbf{(c)} $Q$-factor measurement for $\nu_\textrm{p}^{\textrm{cav}}=c/\lambda_\textrm{p}^{\textrm{cav}}=473.233\,\textrm{THz}$. Sidebands created by an EOM at $\nu_\textrm{pump}\pm3.9\,\textrm{GHz}$ act as a frequency ruler to extract the cavity linewidth.
\textbf{(d)} $Q$-factor as a function of cavity resonance frequency.
}
\label{fig:Fig_2}
\end{figure}

Next, we verify that we are able to establish the double resonance condition by coupling an additional laser resonant with mode $q_\textrm{eff}=17$ at ${\lambda_\textrm{pump}=632.99\,\textrm{nm}}$ into the cavity. The resulting Raman scattered light is at a wavelength of $\lambda_{\textrm{R}}$=(1/$\lambda_{\textrm{pump}}-\Delta\nu_\textrm{R}/c)^{-1}=692.25\,\textrm{nm}$, where $c$ is the speed of light.
We then tune the mirror separation and record spectra from the cavity (Fig.\,\ref{fig:Fig_2}\,(a)).
As expected, the cavity modes with effective mode numbers $q_\textrm{eff}=16...18$ and wavelengths in the range of $\lambda=670...700\,\textrm{nm}$ redshift with increasing mirror separation.
When the pump laser at $\lambda_\textrm{pump}=632.99\,\textrm{nm}$ is resonant with the $q_\textrm{eff}=18...20$ modes for relative mirror separations of $-316.5\,\textrm{nm}$, $0\,\textrm{nm}$ and $+316.5\,\textrm{nm}$, narrow peaks appear in the spectrum at CCD pixels corresponding to $\lambda=691.19\,\textrm{nm}$ and $\lambda=691.32\,\textrm{nm}$, as highlighted in the insets of Fig.\,\ref{fig:Fig_2}\,(a).

 The linecut at $\lambda=691.19\,\textrm{nm}$ clearly shows that the cavity resonances for $q_\textrm{eff}=16$ and $q_\textrm{eff}=18$ appear at smaller ($-345.5\,\textrm{nm}$) and larger mirror separations ($+345.5\,\textrm{nm}$) than the Raman peaks, respectively. Only for a mirror separation of $t_\textrm{a}=4.18\,\upmu\textrm{m}$ $\lambda_\textrm{pump}$ and $\lambda_\textrm{R}$ are, within the spectrometer resolution, simultaneously resonant with the cavity for $q_\textrm{eff}=19$ and $q_\textrm{eff}=17$. For this double resonance condition the signal intensity is increased by over three orders of magnitude compared to the other peaks.

In the following we denote the wavelength of the cavity mode with $q_\textrm{eff}=17$ close in wavelength to that of the pump ($\lambda_\textrm{pump}$) as $\lambda_\textrm{p}^\textrm{cav}$; and the wavelength of the cavity mode with $q_\textrm{eff}=19$ close in wavelength to that of the Raman photon ($\lambda_{\textrm{R}}$) as $\lambda_\textrm{S}^\textrm{cav}$. An analogous notation is adapted for the corresponding frequencies $\nu$.

A faster way to confirm that the double resonance condition is satisfied is displayed in Fig.\,\ref{fig:Fig_2}\,(b). Here, we only couple the diode laser at $\lambda_\textrm{pump}$ into the cavity and record the cavity transmission using a photodiode located beneath the bottom mirror. The transmission spectrum reveals several peaks at mirror separations where the pump laser is resonant with the cavity. These peaks are associated with fundamental and higher order cavity modes. Simultaneously, we measure the cavity emission at wavelengths $>644\textrm{nm}$ using a single photon counting module. A strong signal is observed only when $\lambda_\textrm{pump}$ is resonant with mode $q_\textrm{eff}=19$, while at the same time $\lambda_\textrm{R}$ is resonant with $q_\textrm{eff}=17$.
The correlation between a peak in transmission (signifying a resonant pump laser) and a strong peak in cavity emission at longer wavelengths (signifying a resonant Raman process) is a clear demonstration that the double resonance condition is satisfied. 

%%%%%%%%%%%%%%%%%%%%%%%%%%%%%%%%%%%%%%%%%%%%%%%%%%%%%%%%%%%%%%%%%%%%%%%%%%%
% FIG. 2
%%%%%%%%%%%%%%%%%%%%%%%%%%%%%%%%%%%%%%%%%%%%%%%%%%%%%%%%%%%%%%%%%%%%%%%%%%%
%%%%%%%%%%%%%%%%%%%%%%%%%%%%%%%%%%%%%%%%%%%%%%%%%%%%%%%%%%%%%%%%%%%%%%%%%%%

Next, we determine the quality factor of the pump mode of the cavity, $Q_\textrm{p}$, following the method reported in Ref.\,\cite{Flagan2021}. To extract the cavity linewidth, we keep the laser frequency $\nu_\textrm{pump}$ fixed while scanning the cavity length, monitoring the reflected light on a photodiode. An electro-optic modulator (EOM, Jenoptik PM635) is used to create laser side-bands at $\nu_\textrm{pump}\pm 3.9\,\textrm{GHz}$, thereby providing a frequency ruler to extract the cavity linewidth. Fig.\ref{fig:Fig_2}\,(c) shows the reflected signal averaged over 200 scans for $\nu_\textrm{pump}=473.233\,\textrm{THz}$. Assuming a linear response of the piezo across the resonance, we extract a cavity mode full width at half maximum (FWHM) linewidth of $\delta\nu_\textrm{p}^\textrm{cav}= (1.59\pm0.05)\,\textrm{GHz}$ corresponding to a quality factor of $Q_\textrm{p}=297\,000\pm500$. Fig.\,\ref{fig:Fig_2}\,(d) shows the dependence of $Q_\textrm{p}$ on the cavity resonance frequency $\nu_\textrm{p}^\textrm{cav}$. 

%%%%%%%%%%%%%%%%%%%%%%%%%%%%%%%%%%%%%%%%%%%%%%%%%%%%%%%%%%%%%%%%%%%%%%%%%%%
% FIG. 3
%%%%%%%%%%%%%%%%%%%%%%%%%%%%%%%%%%%%%%%%%%%%%%%%%%%%%%%%%%%%%%%%%%%%%%%%%%%
\begin{figure}[t!]
\includegraphics[width=\columnwidth]{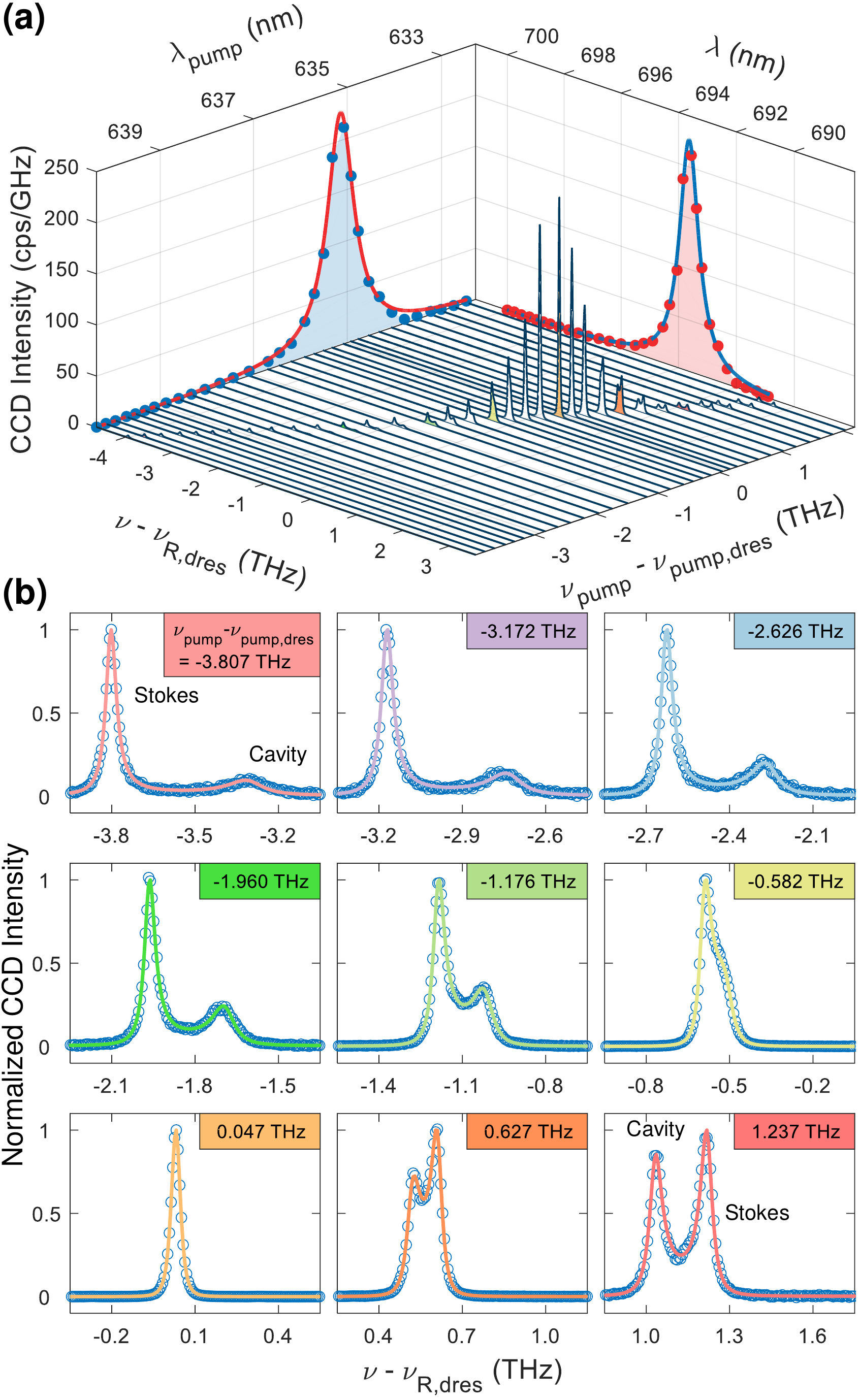}
\caption{\textbf{(a)} Series of cavity spectra for different detunings of the pump laser from the double resonance condition $\nu_\textrm{pump}-\nu_\textrm{pump,dres}$. We ensure that the pump laser is resonant with the cavity for every spectrum by continuously modulating the cavity length. The cavity spectra comprise the Raman signal at $\nu_\textrm{R}$ and the closest cavity mode at $\nu_\textrm{S}^\textrm{cav}$. We fit the amplitude of the Raman signal for every value of $\nu_\textrm{pump}-\nu_\textrm{pump,dres}$ with a Lorentzian (projected blue points) with linewidth of 519.8\,GHz. The center frequency of the fit indicates that the double resonance condition is satisfied when $\nu_\textrm{R,dres}=432.520\,\textrm{THz}$ for $\nu_\textrm{pump,dres}=472.434\,\textrm{THz}$. \textbf{(b)} The individual cavity spectra can be fitted well using a model based on two multiplied Lorentzians centered at the Stokes frequency $\nu_\textrm{R}-\nu_\textrm{R,dres}$ and the frequency of the adjacent cavity mode $\nu_\textrm{S}^\textrm{cav}-\nu_\textrm{R,dres}$. The inset in each panel indicates the pump frequency $\nu_\textrm{pump}-\nu_\textrm{pump,dres}$.
}
\label{fig:Fig_3}
\end{figure}
%%%%%%%%%%%%%%%%%%%%%%%%%%%%%%%%%%%%%%%%%%%%%%%%%%%%%%%%%%%%%%%%%%%%%%%%%%%

%We exploit the fact that the Stokes signal is only generated when the pump laser is resonant with the cavity, $\nu_\textrm{pump}=\nu_\textrm{p}^\textrm{cav}$.
Next, we characterize in more detail the exact detuning dependence of the double resonance condition by tuning the pump laser.
We exploit the fact that Raman signal in the cavity is only generated when a cavity-mode is resonant with the pump laser, $\nu_\textrm{pump}=\nu_\textrm{p}^\textrm{cav}$. In this event, two signals will appear in the spectra: Raman scattering of the pump laser in the cavity at $\nu_\textrm{R}=\nu_\textrm{pump}+\Delta\nu_\textrm{R}$, and the cavity mode nearest in frequency to this Raman signal, $\nu_\textrm{S}^\textrm{cav}$, which is fed by PL light that is non-resonantly generated by the pump laser.
Using this technique it is not necessary to keep the cavity at one particular resonance -- this circumvents any problems caused either by drift or acoustic and thermal noise.

We modulate the cavity length continuously and record the resulting cavity spectra (Fig.\,\ref{fig:Fig_3}\,(a)) varying $\nu_\textrm{pump}=\nu_\textrm{p}^\textrm{cav}$ ($\lambda_\textrm{pump}=\lambda_\textrm{p}^\textrm{cav}$) from 468.475\,THz (639.932\,nm) to 474.471\,THz (631.845\,nm). We find that for the pump frequency $\nu_\textrm{pump,dres}=472.434\,\textrm{THz}$ the double resonance condition is fulfilled, $\nu_\textrm{R}=\nu_\textrm{S}^\textrm{cav}=\nu_\textrm{R,dres}=432.520\,\textrm{THz}$. These values are different from those in Fig.\,\ref{fig:Fig_2} due to a slightly different lateral position of the cavity mode corresponding to a different diamond thickness.
%The center of this Lorentzian corresponds to the pump frequency satisfying the double-resonance condition $\nu_\textrm{pump,dres}=\nu_\textrm{p,dres}=472.434\,\textrm{THz}$ (634.570\,nm). The corresponding Raman frequency is $\nu_\textrm{R,dres}=\nu_\textrm{S,dres}=432.520\,\textrm{THz}$ (693.129\,nm).
 We determine a Raman shift of $\Delta\tilde{\nu}_\textrm{R}=\nu_\textrm{pump,dres}/c-\nu_\textrm{R,dres}/c=1331.4\,\textrm{cm}^{-1}$ ($\Delta\nu_\textrm{R}=c\Delta\tilde{\nu}_\textrm{R}=39.914\,\textrm{THz}$), in good agreement with the previously reported value, $\sim1332\,\textrm{cm}^{-1}$\,\cite{Ferrari2000,Prawer2004,Mildren2013}.
 We plot the peak Raman counts for different detunings of the pump laser from the double resonance condition, $\nu_\textrm{pump}-\nu_\textrm{pump,dres}$ (projected blue points in Fig.\,\ref{fig:Fig_3}\,(a)). We find that these peak counts follow a Lorentzian with FWHM linewidth of 519.8\,GHz\,\cite{Petrak2014}. The corresponding projected Raman amplitude is fitted well by a Lorentzian with FWHM linewidth of 502.9\,GHz (projected red points in Fig.\,\ref{fig:Fig_3}\,(a)).

\begin{figure}[t]
 \includegraphics[width=\columnwidth]{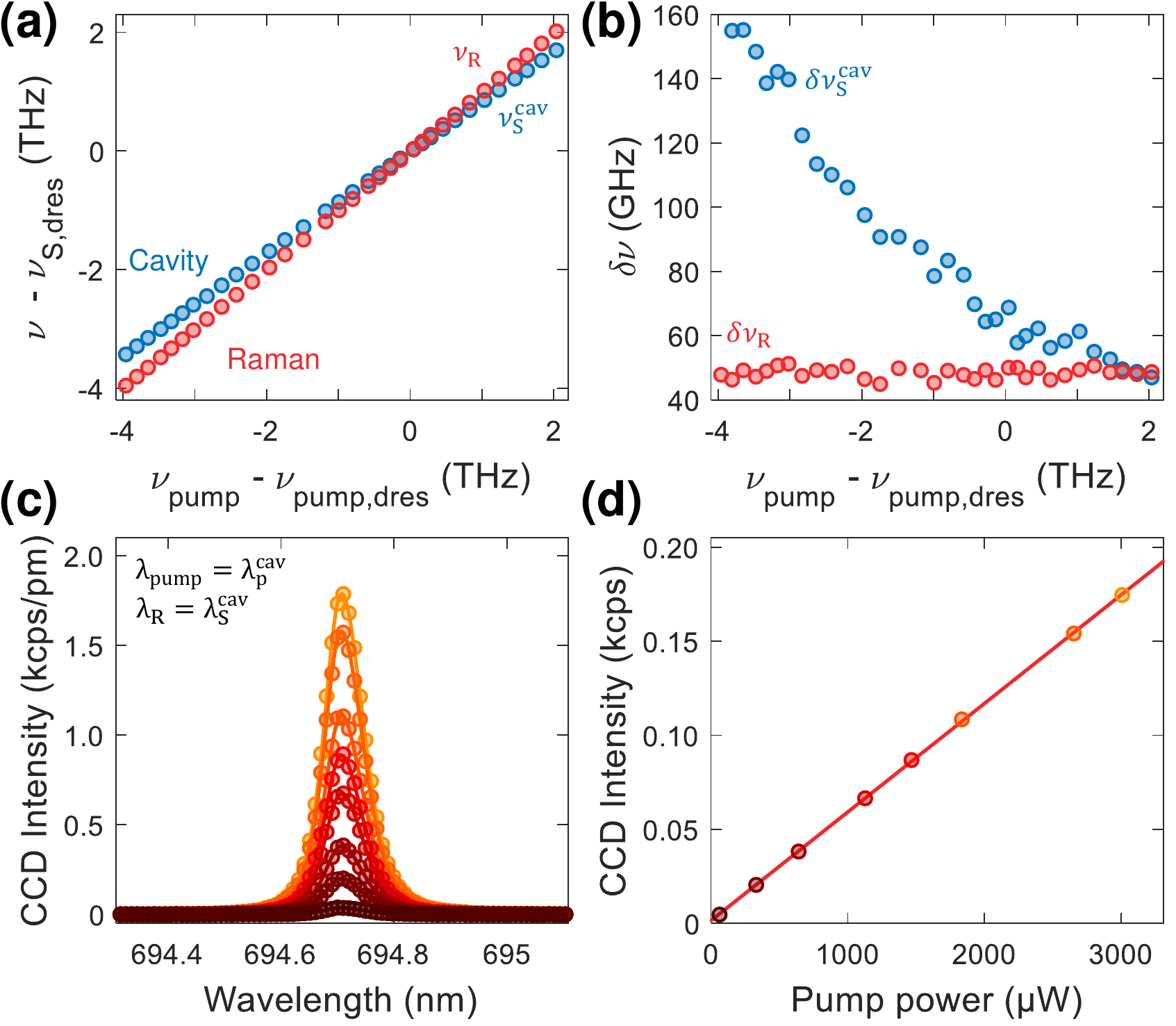}
\caption{\textbf{(a)} Peak position of the cavity ($\nu_\textrm{S}^\textrm{cav}$) and Raman scattered light ($\nu_\textrm{R}$) as a function of pump frequency $\nu_\textrm{pump}$ relative to the double resonance condition at $\nu_\textrm{pump,dres}$ and $\nu_\textrm{R,dres}$.
\textbf{(b)} Linewidth of the cavity ($\delta\nu_\textrm{S}^\textrm{cav}$) and the Raman gain ($\delta\nu_\textrm{R}$) versus pump frequency $\nu_\textrm{pump}-\nu_\textrm{pump,dres}$.
\textbf{(c)} Spectrally resolved cavity signal with increased pump power at the double-resonance condition.
\textbf{(d)} Integrated peak intensity of peaks in (c) as a function of pump power. The linear behavior suggests that no lasing occurs.}
\label{fig:Fig_4}
\end{figure}

We fit the individual spectra for different detunings of the pump laser (with respect to the double-resonance frequency $\nu_{\textrm{pump,dres}}$) to the product of two Lorentzians describing the cavity mode at $\nu_\textrm{S}^\textrm{cav}$ (FWHM $\delta\nu_\textrm{S}^\textrm{cav}$) and the gain bandwidth of the Raman scattering process at $\nu_\textrm{R}$ (FWHM $\delta\nu_\textrm{R}$). These fits allow the peak positions and linewidths to be extracted (Fig.\,\ref{fig:Fig_3})\,\cite{Riedel2020}. Figures\,\ref{fig:Fig_4}\,(a,b) display the results of these fits. Over the tuning range of the pump laser, the detuning between $\nu_\textrm{S}^\textrm{cav}$ and $\nu_\textrm{R}$ varies from $-319.7\,\textrm{GHz}$ to $526.7\,\textrm{GHz}$. The linewidth of the Raman gain for the different fits is $\delta\nu_\textrm{R}=(48.3\pm1.6)$\,GHz corresponding to $Q_\textrm{R}=8\,960\pm290$. This Raman linewidth agrees well with previously reported values (40.8...47.8\,GHz)\,\cite{Anderson2018,Lee2012,Riedel2020}, indicating low strain in the diamond membrane. The linewidth of the cavity mode closest to $\nu_\textrm{R}$ at $\nu_\textrm{S}^\textrm{cav}$ decreases from $\delta\nu_\textrm{S}^\textrm{cav}=(167.3\pm0.8)\,\textrm{GHz}$ to $(47.0\pm0.4)\,\textrm{GHz}$ for increasing $\nu_\textrm{pump}=\nu_\textrm{p}^\textrm{cav}$, which is expected from the increase in reflectivity on approaching the stopband center of the DBR mirror coatings. The corresponding $Q$-factor increases from $Q_\textrm{p}=2\,570\pm90$ to $9\,250\pm90$. At the double-resonance condition, the $Q$-factor of the Stokes cavity mode is $Q_\textrm{S,dres}=6\,650\pm50$.

%%%%%%%%%%%%%%%%%%%%%%%%%%%%%%%%%%%%%%%%%%%%%%%%%%%%%%%%%%%%%%%%%%%%%%%%%%%
% FIG. 4
%%%%%%%%%%%%%%%%%%%%%%%%%%%%%%%%%%%%%%%%%%%%%%%%%%%%%%%%%%%%%%%%%%%%%%%%%%%
%%%%%%%%%%%%%%%%%%%%%%%%%%%%%%%%%%%%%%%%%%%%%%%%%%%%%%%%%%%%%%%%%%%%%%%%%%%

Next, we perform double-resonance measurements for different pump powers $P_\textrm{pump}$ (as measured before the sample objective) (Fig.\,\ref{fig:Fig_4}\,(c,d)). Up to the largest available pump power in the experiment, the intensity increases linearly: there is no superlinear dependence presaging Raman lasing.

To estimate the threshold power required to establish Raman lasing, we analyze the performance of the cavity. Raman scattering in a microcavity can be described using classical coupled mode equations\,\cite{Yang2007,Checoury2010a}. Lasing occurs when the round-trip gain equals the round-trip loss. Assuming that both the pump laser and the Raman light are resonant with the cavity, $\lambda_\textrm{pump}=\lambda_\textrm{p}^\textrm{cav}$ and $\lambda_\textrm{R}=\lambda_\textrm{S}^\textrm{cav}$, the lasing threshold $P_\textrm{th}$ can be calculated via (see Appendix\,\ref{sec:thres}):
\begin{equation}
P_\textrm{th} = \frac{1}{\eta}
\frac{2n_\textrm{S}n_\textrm{p}\pi^2}{\lambda_\textrm{S}^\textrm{cav}\lambda_\textrm{p}^\textrm{cav}g_\textrm{R}^\textrm{B}}
\frac{V_\textrm{R}(Q_\textrm{S}+Q_\textrm{R})}{Q_\textrm{S}^2Q_\textrm{p}}\,.
\label{eq:threshold}
\end{equation}
Here, $\lambda_\textrm{p(S)}^{\textrm{cav}}$, $n_\textrm{p(S)}$ and $Q_\textrm{p(S)}$ are the wavelengths, refractive indices and $Q$-factors for the cavity modes resonant with the pump laser and the Raman light, respectively. $Q_\textrm{R}$ is the quality factor corresponding to the bandwidth of the Raman gain. The bulk Raman gain coefficient in the employed pump wavelength range is $g_\textrm{R}^\textrm{B}\sim40\,\textrm{cm/GW}$\,\cite{Savitski2013}. The power incoupling efficiency $\eta$ can be extracted from the cavity transmission measurement displayed in Fig.\,\ref{fig:Fig_2}\,(c)\,\cite{Nagourney2014,Gallego2016}. From the dip in reflection, we infer a power incoupling efficiency of $\eta_\text{C}=1-P_\textrm{R}/P_\textrm{0}=0.45$. 

Modeling the cavity using a one-dimensional transfer-matrix model along with Gaussian optics, we estimate a Raman mode volume of $V_\textrm{R}=109.85\,\upmu\textrm{m}^3$ (see Appendix\,\ref{sec:VR}). Taking $n_\textrm{p(S)}=n_\textrm{dia}=2.4$, we find $P_\textrm{th}=187.50\,\textrm{mW}$. This relatively low threshold power constitutes a reduction in threshold power by more than an order of magnitude with respect to a bulk Raman laser in the visible\,\cite{Spence2010,Yang2020,Kitzler2017}
%, and is comparable to the threshold reported for integrated resonators in the near infrared\,\cite{Latawiec2015,Latawiec2018}.
With realistic improvements we predict that our device platform could feature threshold powers in the sub-mW range (see Appendix\,\ref{sec:outlook}).

%%%%%%%%%%%%%%%%%%%%%%%%%%%%%%%%%%%%%%%%%%%%%%%%%%%%%%%%%%%%%%%%%%%%%%%%%%%
% FIG. 5
%%%%%%%%%%%%%%%%%%%%%%%%%%%%%%%%%%%%%%%%%%%%%%%%%%%%%%%%%%%%%%%%%%%%%%%%%%%
\begin{figure}[t]
\includegraphics[width=\columnwidth]{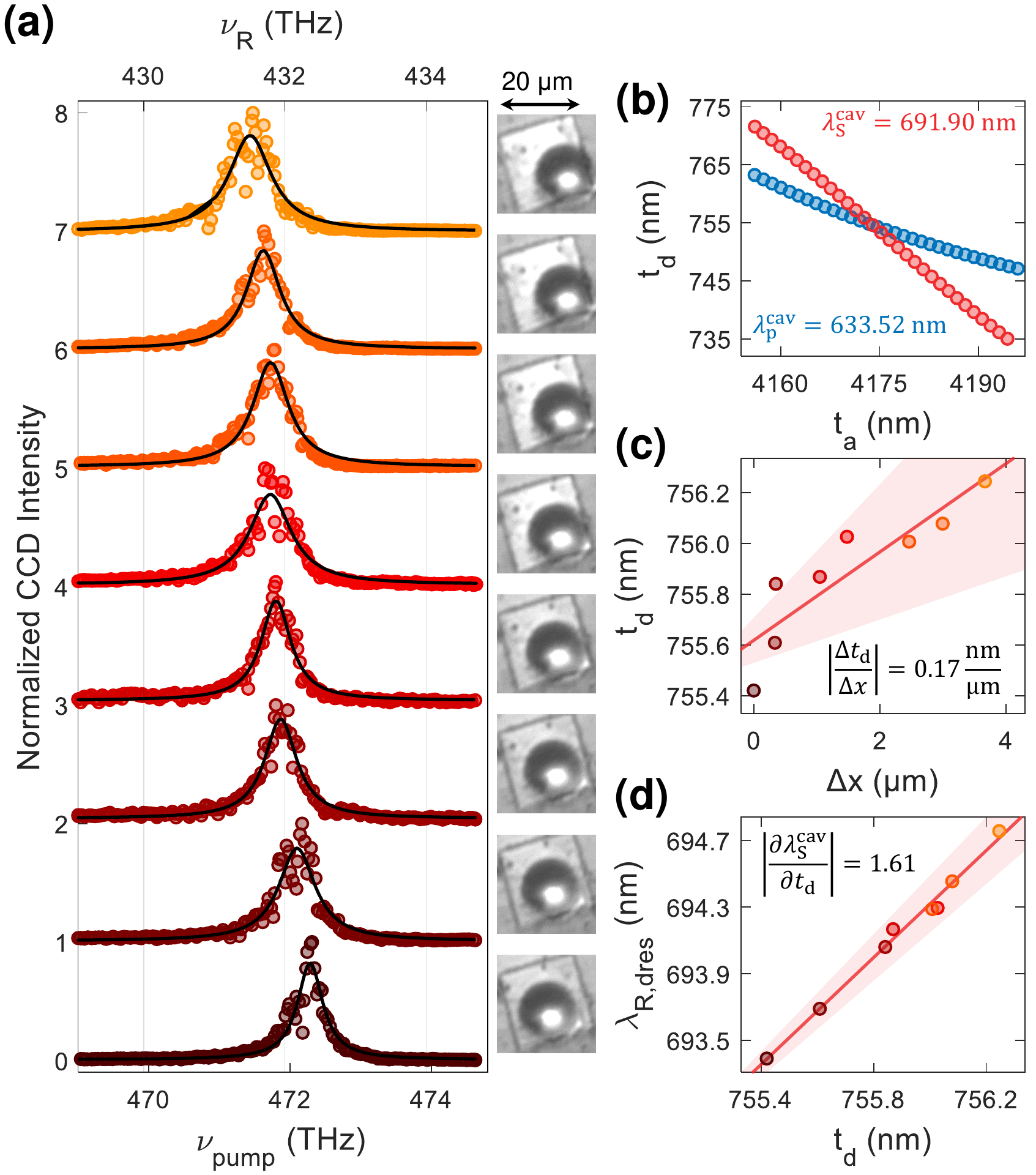}
\caption{\textbf{(a)} Demonstration of the tunability of the double resonance condition. Owing to the thickness gradient, shifting the diamond laterally changes the diamond thickness in the cavity and subsequently the condition for double resonance.
\textbf{(b)} The diamond thickness and the width of the air-gap are extracted from one-dimensional transfer-matrix simulations. The double resonance is condition is satisfied when the pump mode (blue) and the Stokes mode (red) cross. For $\lambda_\textrm{p}^\textrm{cav}=633.52\,\textrm{nm}$ and $\lambda_\textrm{S}^\textrm{cav}=691.90\,\textrm{nm}$, we extract $t_\textrm{d}=754.58\,\textrm{nm}$ and $t_\textrm{a}=4173.98\,\textrm{nm}$.
\textbf{(c)} Lateral displacement of the cavity mode plotted against diamond thickness $t_{\textrm{d}}$. Here, the relative position of the cavity mode is calculated with respect to the corners of the diamond. Extracting the diamond thickness from the double-resonance measurements in (a) gives a thickness gradient of $0.17\,\textrm{nm}/\upmu\textrm{m}$.
\textbf{(d)} Linear shift of the double-resonance condition with the diamond thickness resulting in a red-shift of the Stokes wavelength.}
\label{fig:Fig_5}
\end{figure}
%%%%%%%%%%%%%%%%%%%%%%%%%%%%%%%%%%%%%%%%%%%%%%%%%%%%%%%%%%%%%%%%%%%%%%%%%%%

Next, we demonstrate the possibility to tune the double resonance condition by changing the thickness of the diamond layer within the cavity mode \textit{in situ} (Fig.\,\ref{fig:Fig_5}\,(a)). To this end, we laterally displace the cavity mode with respect to the diamond membrane, exploiting a small thickness gradient (Fig.\,\ref{fig:Fig_1}\,(a)). Over the lateral fine-tuning range of the nanopositioner (travel range $\sim4\,\upmu\textrm{m}$), the double resonance condition can be tuned from $\nu_\textrm{pump,dres}=471.44\,\textrm{THz}$ to $472.29\,\textrm{THz}$ ($\nu_\textrm{R,dres}=431.52\,\textrm{THz}$ to $432.37\,\textrm{THz}$), a continuous tuning range of 0.85\,THz. Considering that the width of the double resonance gain profile is $>$500\,GHz, this would enable a $>$THz continuous tuning range of the lasing frequency.
 
To extract the exact diamond thickness, we perform one-dimensional transfer-matrix-based simulations of the cavity mode structure (see Appendix\,\ref{sec:sim}). For these simulations, we use the exact mirror structure obtained from fitting the mirror stopband (Fig.\,\ref{fig:Fig_1}\,(c)\,\cite{Riedel2020,Flagan2021}), and sweep the width of the air-gap $t_\textrm{a}$ and the diamond thickness $t_\textrm{d}$ for fixed wavelengths $\lambda_\textrm{p}^\textrm{cav}$ and $\lambda_\textrm{S}^\textrm{cav}$. The double resonance condition is met whenever the modes for $\lambda_\textrm{p}^\textrm{cav}$ cross the modes for $\lambda_\textrm{S}^\textrm{cav}$. Fig.\,\ref{fig:Fig_5}\,(b) shows a transfer-matrix simulation for $\lambda_\textrm{p}^\textrm{cav}=633.52\,\textrm{nm}$ and $\lambda_\textrm{S}^\textrm{cav}=691.90\,\textrm{nm}$ (extracted from Fig.\,\ref{fig:Fig_4}\,(a)). Here, the two cavity modes overlap for $t_\textrm{a}=4173.98\,\textrm{nm}$ and $t_\textrm{d}=754.85\,\textrm{nm}$.

We extract $t_\textrm{d}$ for all measurements displayed in Fig.\,\ref{fig:Fig_5}\,(a) and plot $t_\textrm{d}$ versus lateral displacement of the cavity mode (Fig.\,\ref{fig:Fig_5}\,(c)). To calibrate the lateral displacement, we use the edges of the diamond ($\sim 18\,\upmu\textrm{m}$) measured with a laser scanning confocal microscope (Keyence Corporation) as a reference. We find a thickness gradient $|\frac{\Delta t_\textrm{d}}{\Delta x}|=\left(0.17\pm0.2\right)\,\textrm{nm}/\upmu\textrm{m}$. % \co{in agreement with the direct laser confocal measurement (?)}.
As shown in Fig.\,\ref{fig:Fig_5}\,(d), we observe a linear shift of $\lambda_\textrm{S}^\textrm{cav}$ with $t_\textrm{d}$. From our simulations, we find that for the right combination of $t_\textrm{d}$ and $t_\textrm{a}$, the double resonance condition can be tuned continuously across the whole mirror stopband corresponding to a continuous tuning range of tens of THz (see. Appendix\,\ref{sec:drestuning}).

%%%%%%%%%%%%%%%%%%%%%%%%%%%%%%%%%%%%%%%%%%%%%%%%%%%%%%%%%%%%%%%%%%%%%%%%%%%
% CONCLUSION
%%%%%%%%%%%%%%%%%%%%%%%%%%%%%%%%%%%%%%%%%%%%%%%%%%%%%%%%%%%%%%%%%%%%%%%%%%%
\section{Conclusion}
In conclusion, we demonstrate a platform for the widely-tunable doubly-resonant enhancement of Raman scattering from diamond based on a tunable open-access microcavity. The \textit{in situ} tuning capability of our device provides a convenient way to establish a double resonance condition in which both pump and Raman wavelengths are resonant with a cavity mode. Exploiting a slight thickness gradient of the incorporated diamond membrane enables the doubly-resonant configuration to be achieved over a wide tuning range of more than 1\,THz. These results, together with the high quality factors of the cavity in the visible wavelength range, suggest that Raman lasing can be achieved with the present system. We predict a lasing threshold of $188\,\textrm{mW}$, a reduction by more than an order of magnitude compared to bulk Raman lasers\,\cite{Williams2018}. We anticipate that with realistic improvements of our platform, sub-mW Raman lasing thresholds can be achieved. Importantly, we predict that there are configurations where mode-hop-free tuning of the double resonance condition over tens of THz is possible, in principle limited only by the spectral width of the reflective stopband of the mirrors. These advancements pave the way to a universal, low-power, frequency-shifter. Finally, we note that due to the generic design of our platform, other wide-bandgap Raman laser materials such as aluminum nitride\,\cite{Liu2017a} can readily be incorporated into our device. A wider point is that the integration of materials exhibiting a strong $\chi^{(2)}$ nonlinearity such as silicon carbide\,\cite{Guidry2020,Lukin2020b}, lithium niobate\,\cite{McKenna2021} or gallium phosphide\,\cite{Logan2018,Wilson2020} could enable low-threshold frequency conversion using other nonlinear processes, for instance second-harmonic generation or sum- and difference-frequency mixing.

%%%%%%%%%%%%%%%%%%%%%%%%%%%%%%%%%%%%%%%%%%%%%%%%%%%%%%%%%%%%%%%%%%%%%%%%%%%
% ACKNOWLEDGEMENTS
%%%%%%%%%%%%%%%%%%%%%%%%%%%%%%%%%%%%%%%%%%%%%%%%%%%%%%%%%%%%%%%%%%%%%%%%%%%

\section{Acknowledgments} We acknowledge financial support from NCCR QSIT, a competence center funded by SNF; the Swiss Nanoscience Institute (SNI); ITN networks S$^{3}$NANO and SpinNANO; and SNF Grant No.\ 200021\_143697. DR acknowledges support from the Swiss National Science Foundation (Project P400P2\_194424). 
SF and DR contributed equally to performing the experiments, analyzing the data and development of the theory. 
%DR initiated and conceived the project. All authors prepared the manuscript.

%%%%%%%%%%%%%%%%%%%%%%%%%%%%%%%%%%%%%%%%%%%%%%%%%%%%%%%%%%%%%%%%%%%%%%%%%%%
% APPENDIX
%%%%%%%%%%%%%%%%%%%%%%%%%%%%%%%%%%%%%%%%%%%%%%%%%%%%%%%%%%%%%%%%%%%%%%%%%%%

%\newpage
%%%%%%%%%%%%%%%%%%%%%%%%%%%%%%%%%%%%%%%%%%%%%%%%%%%%%%%%%%%%%%%%%%%%%%%%%%%
% METHODS
%%%%%%%%%%%%%%%%%%%%%%%%%%%%%%%%%%%%%%%%%%%%%%%%%%%%%%%%%%%%%%%%%%%%%%%%%%%
\appendix
\section*{APPENDIX}
\subsection{Methods}
\label{sec:methods}
The core of this experiment is the tunable, planar-concave Fabry-Perot microcavity\,\cite{Greuter2014,Barbour2011} with an embedded diamond micromembrane, depicted schematically in Fig.\,\ref{fig:Fig_1}\,(a). The microcavity comprises two fused silica substrates exhibiting highly reflective dielectric mirror coatings (ECI evapcoat). Prior to applying the coating, we fabricate an array of spherical micro-indentations via CO$_2$ laser ablation\,\cite{Hunger2012} in one of the substrates. The micro-indentations feature small radii of curvature ${R_{\textrm{cav}}\sim10\,\upmu\textrm{m}}$ and depths $d\sim 1.5\,\upmu\textrm{m}$. We employ 14 (15) $\lambda_{\textrm{c}}/4$ layers of a SiO$_2$/Ta$_2$O$_5$ distributed Bragg reflector for the curved top (planar bottom) mirrors. From a white-light transmission measurement\,\cite{Riedel2020,Flagan2021,Najer2021}, the center of the stopband of the top mirror is determined to be $\lambda_{\textrm{c}}=625\,\textrm{nm}$ (Fig.\,\ref{fig:Fig_1}\,(c)). Using a transfer-matrix-based refinement algorithm (Essential Macleod) we can reconstruct the reflection spectrum utilizing an individual layer-thickness tolerance of 3\% with $n_{\textrm{SiO}_2}=1.46$ and $n_{\textrm{Ta}_2\textrm{O}_5}=2.11$. Using the same approach we find $\lambda_{\textrm{c}}=629\,\textrm{nm}$ for the top mirror.
 
Starting with commercially available high-purity, $\langle$100$\rangle$-cut single crystal diamond (Element Six), we fabricate membranes via inductively-coupled reactive-ion etching and electron-beam lithography\,\cite{Maletinsky2012, Riedel2014, Appel2016}. We then transfer membranes with typical dimensions $\sim20\times20\times0.8\,\upmu\textrm{m}^3$ to the planar mirror using a micromanipulator\,\cite{Riedel2014, Riedel2017}. The diamond membranes exhibit a slight thickness gradient introduced during the thinning of the diamond\,\cite{Challier2018}. 

The bottom mirror is mounted on a stack of xyz-piezoelectric nanopositioners (attocube, 2$\times$ANPx51 and ANPz51) and placed inside a homebuilt titanium ``cage"; the top mirror is rigidly attached to the top of the cage\,\cite{Flagan2021}. By applying a voltage to the nanopositioners, the bottom mirror can be moved in all three dimensions with respect to the top mirror, offering both spatial and spectral tunability\,\cite{Riedel2017}. Finally, the titanium cage is mounted on top of a high-precision mechanical stage (Newport, M-562-XYZ) to enable the cavity output to be coupled to external detection optics\,\cite{Riedel2020}.

We use a narrow-band tunable red diode-laser as pump laser (Toptica DL Pro 635, $\lambda_\textrm{pump}=630\cdots 640\,\textrm{nm}$, $\delta\nu\leq 500\,\textrm{kHz}$). This pump laser is spectrally filtered (Semrock, FF01-637/7-25 and FF01-650/SP-25) and then coupled into the cavity using an objective of moderate numerical aperture (Microthek, 20$\times$, NA=0.4)\,\cite{Riedel2020}. The Stokes signal is collected via the same objective in a back-scattering geometry (Fig.\,\ref{fig:Fig_1}\,(a)). A combination of a dichroic mirror (cutoff 644\,nm, AHF F48-644) and a long-pass filter (Semrock, BLP02-635R-25) is used to filter the excitation laser from the signal. The Stokes signal is then coupled into a single-mode detection fiber (Thorlabs, 630-HP) and recorded with a spectrometer. 

%%%%%%%%%%%%%%%%%%%%%%%%%%%%%%%%%%%%%%%%%%%%%%%%%%%%%%%%%%%%%%%%%%%%%%%%%%%
% MODE STRUCTURE
%%%%%%%%%%%%%%%%%%%%%%%%%%%%%%%%%%%%%%%%%%%%%%%%%%%%%%%%%%%%%%%%%%%%%%%%%%%
\subsection{Cavity mode structure}
\label{sec:sim}
\begin{figure}[t]
\includegraphics[width=\columnwidth]{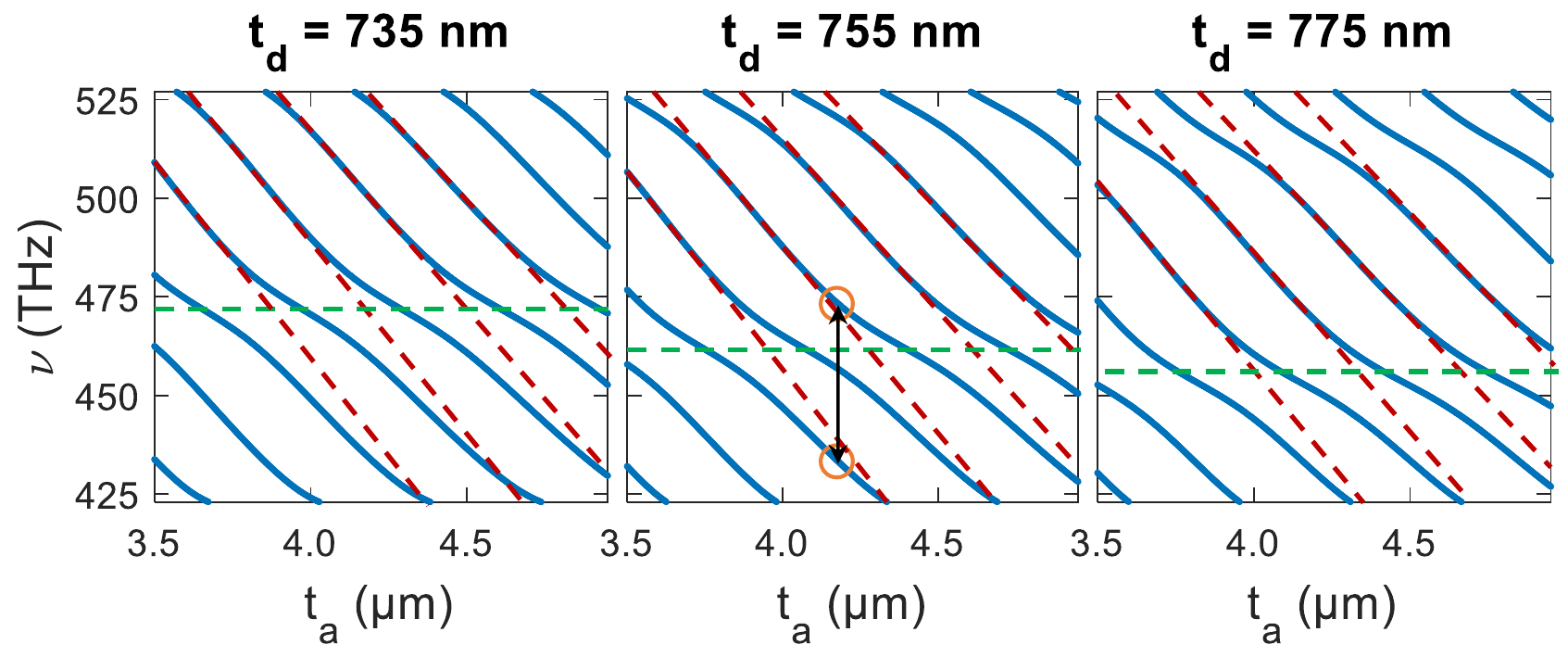}
\caption{One-dimensional transfer-matrix calculation of the cavity mode structure, i. e. resonance frequencies for different mirror separations $t_{\textrm{a}}$, for different diamond thicknesses $t_{\textrm{d}}$. The non-linear mode dispersion arises as a consequence of hybridization of cavity modes resonant with the air-gap (indicated by the dashed burgundy line) and the diamond (dashed green line), respectively. The hybridization leads to avoided crossings whose position depends on the exact diamond thickness. The central panel shows the mode-structure for the extracted diamond thickness ($t_{\textrm{d}}\simeq755\,\textrm{nm}$) from Fig.\,\ref{fig:Fig_5}\,(b). The orange circles indicate a pair of cavity modes whose frequencies are separated by the Raman shift of $\Delta\nu_\textrm{R}\sim40\,\textrm{THz}$ (black arrow).}
\label{fig:Mode_Struct}
\end{figure}

%%%%%%%%%%%%%%%%%%%%%%%%%%%%%%%%%%%%%%%%%%%%%%%%%%%%%%%%%%%%%%%%%%%%%%%%%%%
Conceptually, the cavity mode can be described using a coupled two-cavity model: one cavity is confined to the diamond bound by the bottom DBR and the diamond-air interface; the other cavity is confined to the air bounded by the diamond-air interface and the top DBR\,\cite{Riedel2020,Flagan2021,Janitz2020}. Figure\,\ref{fig:Mode_Struct} displays a one-dimensional transfer-matrix calculation of the cavity mode structure, using the mirror structure extracted from Fig.\,\ref{fig:Fig_1}\,(c). These calculations confirm that the locations of the avoided crossings in the mode structure depend on the choice of diamond thickness $t_{\textrm{d}}$. We find that for $t_{\textrm{a}}=4173.98\,\textrm{nm}$ and $t_{\textrm{d}}=754.58\,\textrm{nm}$, the spectrum observed in Fig.\,\ref{fig:Fig_1}\,(c) is reproduced well.

%%%%%%%%%%%%%%%%%%%%%%%%%%%%%%%%%%%%%%%%%%%%%%%%%%%%%%%%%%%%%%%%%%%%%%%%%%%
% FIGURE 6 - MODE STRUCTURE
%%%%%%%%%%%%%%%%%%%%%%%%%%%%%%%%%%%%%%%%%%%%%%%%%%%%%%%%%%%%%%%%%%%%%%%%%%%

%%%%%%%%%%%%%%%%%%%%%%%%%%%%%%%%%%%%%%%%%%%%%%%%%%%%%%%%%%%%%%%%%%%%%%%%%%%
% TUNING OF DOUBLE RESONANCE CONDITION
%%%%%%%%%%%%%%%%%%%%%%%%%%%%%%%%%%%%%%%%%%%%%%%%%%%%%%%%%%%%%%%%%%%%%%%%%%%
\subsection{Tuning of double resonance condition}
\label{sec:drestuning}
Open-access microcavities offer a convenient tuning mechanism of their resonance frequency simply by changing the separation of the two mirrors ($t_{\textrm{a}}$) using a piezoelectric nanopositioner. Importantly, such cavities offer another tuning mechanism where, rather than the width of the air-gap, the thickness of the material layer is changed \textit{in situ}. Here, a small thickness gradient in the diamond membrane converts a lateral displacement of the cavity mode into a change of the membrane thickness ($t_\textrm{d}$). Tuning both $t_{\textrm{a}}$ and $t_{\textrm{d}}$ allows both the absolute wavelength and the spacing of the cavity modes to be controlled. As a consequence, a gradient in the diamond thickness $|\frac{\Delta t_\textrm{d}}{\Delta x}|$ enables the double resonance condition to be satisfied for different pairs of wavelengths. In Fig.\,\ref{fig:Fig_5}\,(a) we demonstrate experimentally a continuous tuning range of the double resonance condition by 0.85\,THz. This is achieved by changing the diamond thickness by $\sim0.6\,\textrm{nm}$, from 755.5\,nm to 756.2\,nm.

\begin{figure}[bt]
\includegraphics[width=\columnwidth]{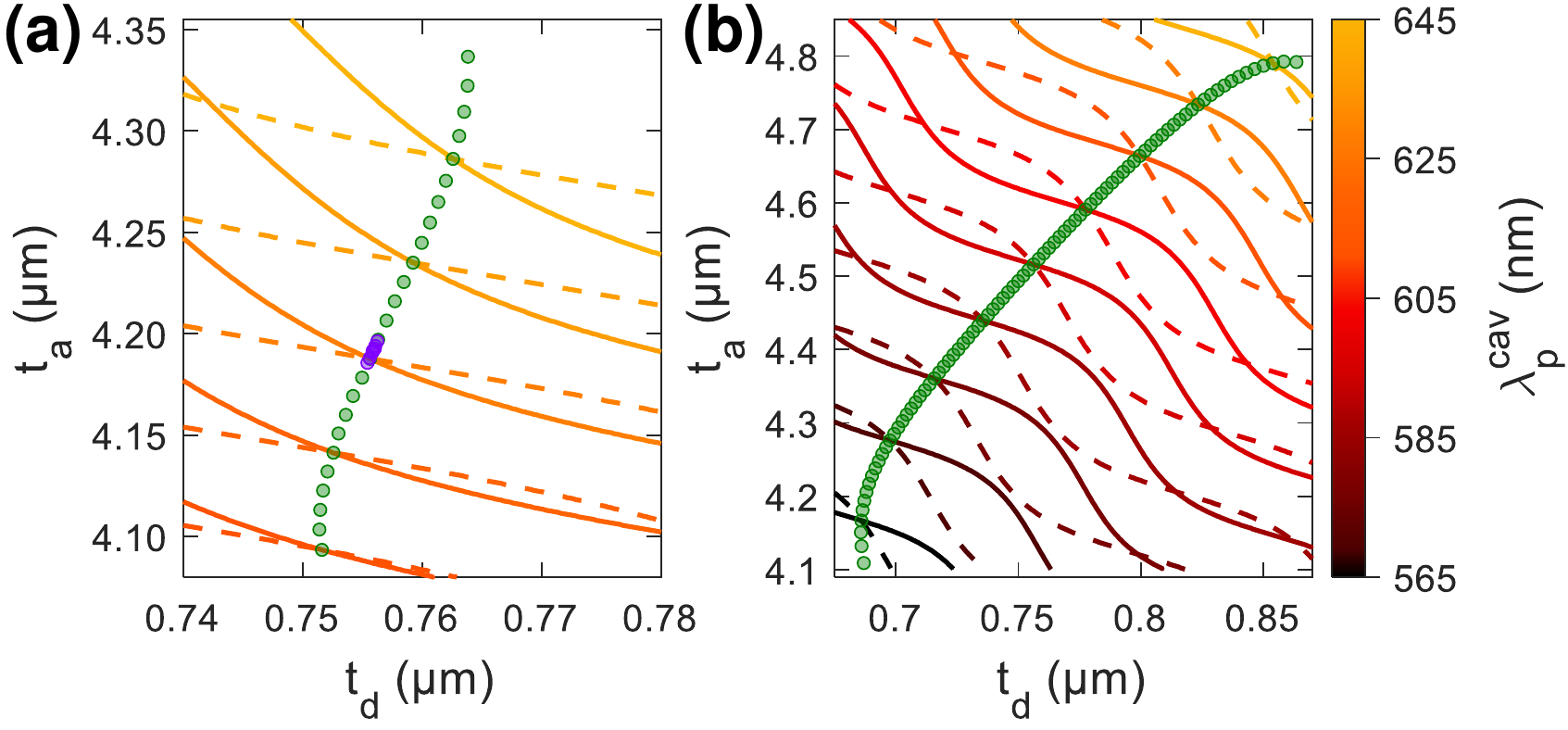}
\caption{One-dimensional transfer-matrix calculations of the cavity resonances as a function of $t_{\textrm{a,d}}$ for different combinations of $\lambda_\textrm{p}^\textrm{cav}$ (solid line) and $\lambda_\textrm{S}^\textrm{cav}$=(1/$\lambda_\textrm{p}^\textrm{cav}-\Delta\nu_\textrm{R}/c)^{-1}$ (dashed line) at specific wavelengths. The double resonance condition is satisfied when the corresponding solid and dashed lines cross (indicated by same color). By changing the diamond thickness, the condition for the double-resonance can be tuned continuously, as indicated by the green circles. \textbf{(a)} Calculations for pairs of $t_{\textrm{a,d}}$ in a range accessible with the device presented in the main text.
The calculations suggest a continuous tuning range of $17.3\,\textrm{THz}$. The experimentally verified tuning is indicated by the purple circles. \textbf{(b)} Optimizing the choice of $t_{\textrm{a,d}}$ enables continuous tuning across the entire reflective stopband, amounting to $72.2\,\textrm{THz}$. For simplicity, only cavity modes for which continuous tuning is possible are included in (b).
} 
\label{fig:tuningdres}
\end{figure}

To explore this tuning mechanism in more detail, we perform one-dimensional transfer-matrix calculations (Essential Macleod). We calculate the combinations of air-gap width $t_{\textrm{a}}$ and diamond thickness $t_{\textrm{d}}$ at which specific wavelengths are resonant.
We perform these calculations for cavity wavelengths within the tuning range of our pump laser, i.e. $\lambda_\textrm{p}^\textrm{cav}=630\cdots 640\,\textrm{nm}$ and for a range of $t_{\textrm{a}}$ and $t_{\textrm{d}}$ accessible with the device presented in the main text (solid lines Fig.\,\ref{fig:tuningdres}\,(a)). 
We then add calculations for the corresponding wavelengths red-shifted by the Raman shift at $\lambda_\textrm{S}^\textrm{cav}=(1/\lambda_\textrm{p}^\textrm{cav}-\Delta\nu_\textrm{R}/c)^{-1}$ (dashed lines Fig.\,\ref{fig:tuningdres}\,(a)).
At pairs of $t_{\textrm{a}}$ and $t_{\textrm{d}}$ where the solid and dashed line cross, the double-resonance condition is satisfied.

 We find that, in principle, the double resonance condition can be tuned continuously from $\lambda_\textrm{pump,dres}=625.00\,\textrm{nm} (\lambda_\textrm{R,dres}=681.75\,\textrm{nm})$ to $\lambda_\textrm{pump,dres}=649.00\,\textrm{nm} (\lambda_\textrm{R,dres}=710.41\,\textrm{nm})$ ($17.3\,\textrm{THz}$) by changing the diamond thickness from 751.4\,nm to 763.8\,nm (green points in Fig.\,\ref{fig:tuningdres}\,(a)). The experimentally demonstrated tuning range is indicated by the purple points in Fig.\,\ref{fig:tuningdres}\,(a).

By optimizing the choice of $t_{\textrm{a,d}}$, we find a configuration which in principle allows the double-resonance condition to be tuned in a mode-hope free fashion harnessing the whole stopband of the mirror (72.2\,THz, $\lambda_\textrm{pump,dres}=565\cdots 645\,\textrm{nm}$, $\lambda_\textrm{R,dres}=610.98\cdots 705.62\,\textrm{nm}$, see Fig.\,\ref{fig:tuningdres}\,(b)).

%%%%%%%%%%%%%%%%%%%%%%%%%%%%%%%%%%%%%%%%%%%%%%%%%%%%%%%%%%%%%%%%%%%%%%%%%%%
% FIGURE 7 - TUNING RANGE
%%%%%%%%%%%%%%%%%%%%%%%%%%%%%%%%%%%%%%%%%%%%%%%%%%%%%%%%%%%%%%%%%%%%%%%%%%%
%%%%%%%%%%%%%%%%%%%%%%%%%%%%%%%%%%%%%%%%%%%%%%%%%%%%%%%%%%%%%%%%%%%%%%%%%%%

%%%%%%%%%%%%%%%%%%%%%%%%%%%%%%%%%%%%%%%%%%%%%%%%%%%%%%%%%%%%%%%%%%%%%%%%%%%
% Calculation of Raman mode volume
%%%%%%%%%%%%%%%%%%%%%%%%%%%%%%%%%%%%%%%%%%%%%%%%%%%%%%%%%%%%%%%%%%%%%%%%%%%
\subsection{Calculation of effective Raman mode volume}\label{sec:VR}
We consider a doubly-resonant system ($\omega_\textrm{p}=\omega_\textrm{pump}=\omega_\textrm{p}^\textrm{cav}$ and $\omega_\textrm{S}=\omega_\textrm{R}=\omega_\textrm{S}^\textrm{cav}$). In the following, we omit the "cav" superscripts for concise notation and clarity.
The effective Raman mode volume accounts for the spatial overlap of the pump (p) and Stokes (S) cavity modes and can be determined via\,\cite{Yang2007,Checoury2010}
\begin{equation}
V_{\textrm{R}}=
\frac{\int_\textrm{cav} n^2_{\textrm{p}}(\vec{r})|\vec{E}_{\textrm{p}}(\vec{r})|^2\textrm{d}^3r\times\int_\textrm{cav} n^2_{\textrm{S}}(\vec{r})|\vec{E}_{\textrm{S}}(\vec{r})|^2\textrm{d}^3r}
{\int_{\textrm{dia}}n^2_{\textrm{p}}(\vec{r})|\vec{E}_{\textrm{p}}(\vec{r})|^2\times n^2_{\textrm{S}}(\vec{r})|\vec{E}_{\textrm{S}}(\vec{r})|^2 \textrm{d}^3r}\,,
\label{eq:RamanModeOverlap}
\end{equation}
where $\vec{E}_{p(S)}(\vec{r})$ is the pump (Stokes) electric field at position $\vec{r}$. 
The integrals over the electric field can be calculated following the same approach as reported in Ref.\,\cite{Riedel2020}. 
We approximated the beam waist to be constant $w_{\textrm{0,I}}$ and solve the integral in cylindrical coordinates:
\begin{equation}
\begin{aligned}
&\int_\textrm{cav} n^2(\vec{r})|\vec{E}(\vec{r})|^2\textrm{d}^3r= 
\\
=&\int_{\textrm{cav}}n^2(z)|\vec{E}(z)|^2\textrm{d}z\int_{0}^{2\pi}\textrm{d}\phi\int_{0}^{\infty}re^{-r^2/2w_{\textrm{I}}^2}\textrm{d}r \\
=&2\pi\frac{1}{4}w_{\textrm{I}}^{2}\int_{\textrm{cav}}\epsilon_{\textrm{0}}n^2(z)|\vec{E}(z)|^2\textrm{d}z\,, \label{eq:E_field}
\end{aligned}
\end{equation}
where $\epsilon_{\textrm{R}}=n^2$ and $w_{\textrm{0,I}}$ is the intensity beam-waist given by\,\cite{Flagan2021,vanDam2018}
\begin{equation}
w_{\textrm{0,I}}=\sqrt{\frac{\lambda}{\pi}}\left[\left(t_{\textrm{a}}+\frac{t_{\textrm{d}}}{n_{\textrm{d}}}\right)\times R_{\textrm{cav}}-\left(t_{\textrm{a}}+\frac{t_{\textrm{d}}}{n_{\textrm{d}}}\right)^2\right]^{\frac{1}{4}}\,.
\label{eq:Beam_waist}
\end{equation}

Calculating the respective field profiles according to Eq.\,\ref{eq:E_field} reduces Eq.\,\ref{eq:RamanModeOverlap} to
\begin{equation}
\begin{aligned}
V_{\textrm{R}}&=2\pi\frac{1}{4}
\left(
w_{\textrm{p}}^2+w_{\textrm{S}}^2\right)\times\\
&\frac{\int_\textrm{cav} n^2_{\textrm{p}}(z)|E_{\textrm{p}}(z)|^2\textrm{d}z
\times
\int_\textrm{cav} n^2_{\textrm{S}}(z)|E_{\textrm{S}}(z)|^2dz}
{\int_{\textrm{dia}}n^2_{\textrm{p}}(z)|E_{\textrm{p}}(z)|^2\times n^2_{\textrm{S}}(z)|E_{\textrm{S}}(z)|^2\textrm{d}z} \,.
\end{aligned}
\end{equation}

\begin{figure}[!tb]
\includegraphics[width=\columnwidth]{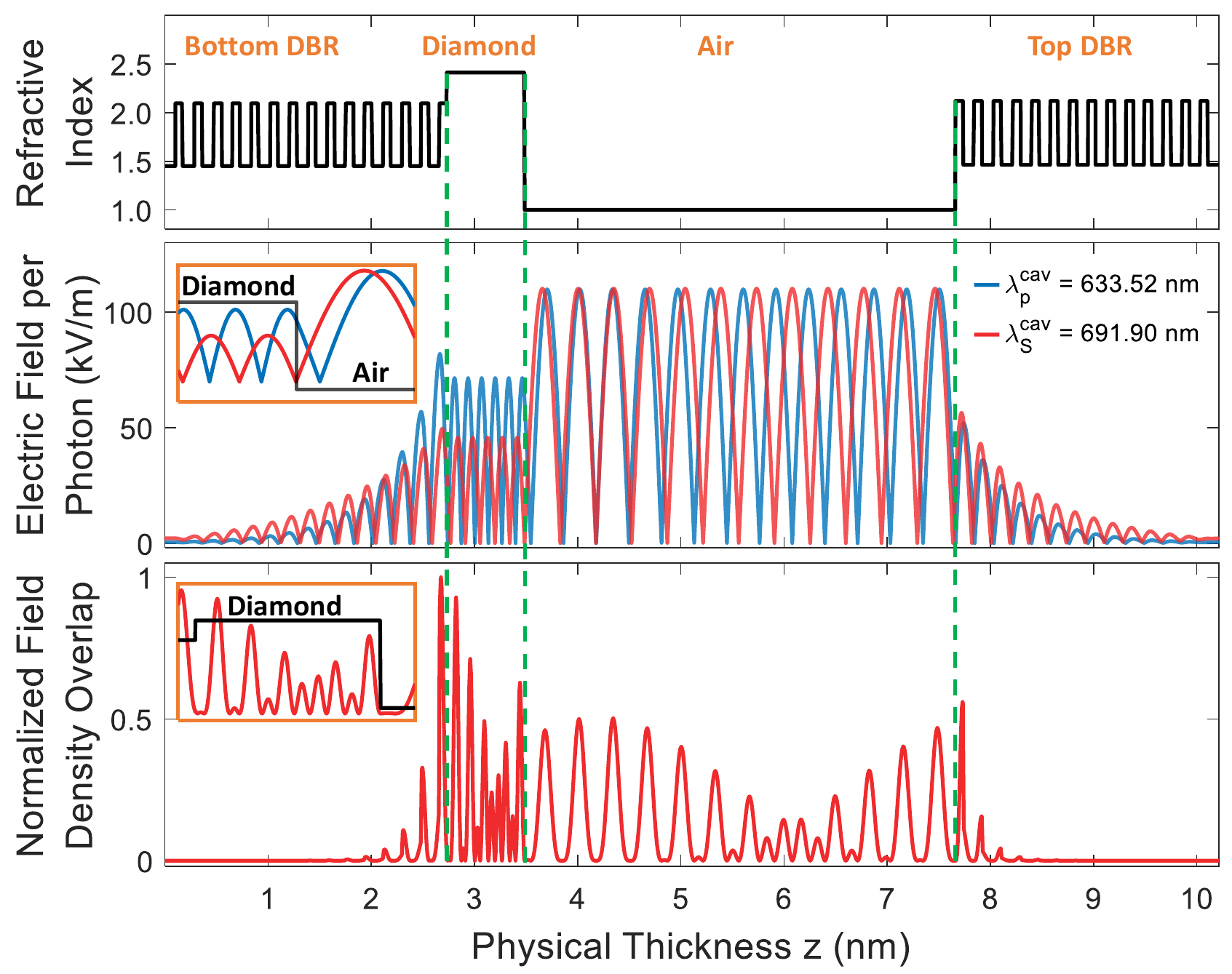}
\caption{One-dimensional transfer-matrix simulation of the cavity for $t_\textrm{a}=4173.98\,\textrm{nm}$ and $t_\textrm{d}=754.58\,\textrm{nm}$. The top panel shows the refractive index profile as a function of cavity length. The middle panel shows the profile of the pump (blue) and Stokes (red) electric field. The bottom panel shows the overlapped energy density calculated according to the denominator in the fraction in Eq.\,\ref{eq:RamanModeOverlap}.
} 
\label{fig:Evac}
\end{figure}

To calculate the Raman mode volume, we approximate the axial vacuum electric-field distribution with a one-dimensional cavity using a transfer-matrix calculation (Essential Macleod).
We use the exact mirror structure obtained from fitting the mirror stopband (Fig.\,\ref{fig:Fig_1}\,(c)) and the combination of $t_{\textrm{a,d}}$ extracted from Fig.\,\ref{fig:Fig_5}\,(b). Fig.\,\ref{fig:Evac} shows the result of our calculations. We determine the electric field profile for the pump and Stokes fields and then their product by numerical integration. Using $R_{\textrm{cav}}=11\,\upmu\textrm{m}$ and $n_{\textrm{p}}\simeq n_{\textrm{S}}=2.4$, we calculate the beam waists according to Eq.\,\ref{eq:Beam_waist}, and find $w_{\textrm{p}} = 1.05\,\upmu\textrm{m}$ and $w_{\textrm{S}} = 1.09\,\upmu\textrm{m}$ taking the $\lambda_\textrm{S,p}^\textrm{cav}$ combination extracted from Fig.\,\ref{fig:Fig_4}\,(a). Finally, we arrive at $V_{\textrm{R}}=109.85\,\upmu\textrm{m}^3$, as quoted in the main text. 

%%%%%%%%%%%%%%%%%%%%%%%%%%%%%%%%%%%%%%%%%%%%%%%%%%%%%%%%%%%%%%%%%%%%%%%%%%%
% FIGURE 8 - Evac
%%%%%%%%%%%%%%%%%%%%%%%%%%%%%%%%%%%%%%%%%%%%%%%%%%%%%%%%%%%%%%%%%%%%%%%%%%%
%%%%%%%%%%%%%%%%%%%%%%%%%%%%%%%%%%%%%%%%%%%%%%%%%%%%%%%%%%%%%%%%%%%%%%%%%%%

%%%%%%%%%%%%%%%%%%%%%%%%%%%%%%%%%%%%%%%%%%%%%%%%%%%%%%%%%%%%%%%%%%%%%%%%%%%
% Calculation of lasing threshold
%%%%%%%%%%%%%%%%%%%%%%%%%%%%%%%%%%%%%%%%%%%%%%%%%%%%%%%%%%%%%%%%%%%%%%%%%%%
\subsection{Calculation of lasing threshold}
\label{sec:thres}

To calculate the lasing threshold we follow the approach presented by Checoury \textit{et al.}\,\cite{Checoury2010a,Boyd1992}. We consider a doubly-resonant system ($\omega_\textrm{p}=\omega_\textrm{pump}=\omega_\textrm{p}^\textrm{cav}$ and $\omega_\textrm{S}=\omega_\textrm{R}=\omega_\textrm{S}^\textrm{cav}$). In the following, we omit the "cav" superscripts for concise notation and clarity. The spacing between the cavity modes is given by $\omega_\textrm{S}=\omega_\textrm{p}-\Delta\omega_\textrm{R}$ and $\Delta\omega_\textrm{R}$ is the Raman shift. The coupled mode equations linking the mean Stokes ($N_\textrm{S}$) and pump photon numbers ($N_\textrm{p}$) are given by:
\begin{equation}
\label{eq:cmpump}
\frac{\textrm{d}N_\textrm{p}}{\textrm{d}t}=
-\frac{N_\textrm{p}}{\tau_\textrm{p}}
-\gamma\frac{N_\textrm{p}}{\tau_\textrm{R}}
-(N_\textrm{S}+1) \frac{N_\textrm{p}}{\tau_\textrm{R}^\textrm{cav}} 
+\kappa_\textrm{in} P_\text{p}
\end{equation}
\begin{equation}
\label{eq:cmStokes}
\frac{\textrm{d}N_\textrm{S}}{\textrm{d}t}=
-\frac{N_\textrm{S}}{\tau_\textrm{S}}
+(N_\textrm{S}+1)\frac{N_\textrm{p}}{\tau_\textrm{R}^\textrm{cav}}\,. 
\end{equation}
Here, $\tau_\textrm{S}=Q_\textrm{S}/\omega_\textrm{S}$ and $\tau_\textrm{p}=Q_\textrm{p}/\omega_\textrm{p}$ are the Stokes and pump photon lifetimes. $\gamma$ describes the Raman scattering into modes other than the cavity mode, and $\tau_\textrm{R}$ is a measure for the spontaneous Raman scattering lifetime in bulk. Stimulated ($N_\textrm{S}$) and spontaneous (+1) Raman scattering into the cavity mode are accelerated with respect to the bulk scattering rate via Purcell enhancement; the corresponding lifetime becomes $\tau_\textrm{R}^\textrm{cav}$. The constant $\kappa_\textrm{in}$ relates the injected pump photon-number per time to the incident pump power $P_\text{p}$.

The spontaneous Raman scattering rate in bulk when the pump mode-polarization is aligned along
the $\langle$110$\rangle$ crystallographic axis can be calculated via\,\cite{Checoury2010a}:
\begin{equation}
\frac{1}{\tau_\textrm{R}}=\frac{2g_\textrm{R}^\textrm{B} c^2 \hbar\omega_\textrm{p}}{3 n_\textrm{p}n_\textrm{S}V} M\,.
\end{equation}
Here, $g_\textrm{R}^\textrm{B}$ denotes the bulk Raman gain and $V$ the mode volume of a hypothetical large cavity. $M$ characterizes the total number of Raman modes into which the system can radiate in such a large cavity with mode volume $V$ for a frequency band of width $\delta\omega_\textrm{R}$\,\cite{Boyd1992}: %p.478 Eq. (10.2.11)
\begin{equation}
M=\frac
{V\omega_\textrm{S}^2 n_\textrm{S}^3}
{2\pi c^3} \delta\omega_\textrm{R}\,.
\end{equation}
$\delta\omega_\textrm{R}$ describes the FWHM linewidth of the gain profile of the Raman scattering process.
Hence:
\begin{equation}
\frac{1}{\tau_\textrm{R}}=\Gamma_\textrm{R}=
\frac{\omega_\textrm{S}^2n_\textrm{S}^2 g_\textrm{R}^\textrm{B}\hbar\omega_\textrm{p}\delta\omega_\textrm{R}}
{3n_\textrm{p}\pi c}\,.
\end{equation}

%In order to calculate the cavity-enhanced Raman decay-rate, we need to account for the spectral overlap of the cavity enhancement with the Raman gain profile. We add a factor of 1/2 to account for the fact that Raman emission occurs along two orthogonal polarizations.
The cavity enhancement is given by a Lorentzian with amplitude $F_\textrm{P}$\,\cite{Barbour2011}. We approximate the Raman gain-profile with a normalized Lorentzian\,\cite{Checoury2010a}. We assume that the cavity is resonant with the Raman scattered light $\omega_\textrm{S}=\omega_\textrm{R}$:

\begin{align}
\frac{1}{\tau_\textrm{R}^\textrm{cav}}=
\frac{\omega_\textrm{S}^2n_\textrm{S}^2 g_\textrm{R}^\textrm{B}\hbar\omega_\textrm{p}\delta\omega_\textrm{R}}
{3n_\textrm{p}\pi c}\int_0^{\infty}\textrm{d}\omega 
& \times \frac{2}{\pi}
\frac{\delta\omega_\textrm{R}}{4(\omega-\omega_\textrm{S})^2+\delta\omega_\textrm{R}^2}
\nonumber \\
&
\times F_\textrm{P}
\frac{\delta\omega_\textrm{S}^2}{4(\omega-\omega_\textrm{S})^2+\delta\omega_\textrm{S}^2}\,.
\end{align}
 For $\omega_\textrm{S}\gg \delta\omega_\textrm{S}$, $\delta\omega_\textrm{R}$ the integrand is close to 0 for $\omega=0$, so the lower limit
of the integral can be extended to negative infinity to obtain an analytical solution:
\begin{equation}
\frac{1}{\tau_\textrm{R}^\textrm{cav}}=\Gamma_\textrm{R}^\textrm{cav}=
\frac{\omega_\textrm{S}^2n_\textrm{S}^2 g_\textrm{R}^\textrm{B}\hbar\omega_\textrm{p}\delta\omega_\textrm{R}}
{3n_\textrm{p}\pi c} \frac{F_\textrm{P}\delta\omega_\textrm{S}}{\delta\omega_\textrm{R}+\delta\omega_\textrm{S}}\,.
\end{equation}

The Purcell enhancement of the system is given by:
\begin{equation}
\frac{\Gamma_\textrm{R}^\textrm{cav}}{\Gamma_\textrm{R}}=F_\textrm{P} \frac{\delta\omega_\textrm{S}}{\delta\omega_\textrm{R}+\delta\omega_\textrm{S}}. 
\end{equation}
This equation resembles the expression for Purcell enhancement of a two-level emitter in a regime in which the linewidth of the cavity and a coupled emitter are comparable\,\cite{Kaupp2016,Romeira2018,Auffeves2010,Meldrum2010}:
\begin{equation}
\frac{\Gamma_\textrm{R}^\textrm{cav}}{\Gamma_\textrm{R}}=\frac{3}{4\pi^2}\left(\frac{\lambda_\textrm{S}^\textrm{cav}}{n_\textrm{S}}\right)^3\frac{1}{V_\textrm{R}}\frac{Q_\textrm{S}Q_\textrm{R}}{Q_\textrm{S}+Q_\textrm{R}}\,.
\end{equation}

The lasing threshold power in the steady state can be calculated from:
\begin{equation}
\Gamma_\textrm{R}^\textrm{cav}N_\textrm{p}(N_\textrm{S}+1) = \frac{N_\textrm{S}}{\tau_\textrm{S}}\,. 
\end{equation}
Using Eq.\,\ref{eq:cmpump} and Eq.\,\ref{eq:cmStokes},
\begin{equation}
\kappa_\textrm{in} P_\textrm{p} = 
\frac{N_\textrm{S}}{\tau_\textrm{S}}
\left(
\frac{1}{\Gamma_\textrm{R}^\textrm{cav}(N_\textrm{S}+1)}
\left(
\frac{1}{\tau_\textrm{p}}
+\frac{\gamma}{\tau_\textrm{R}}
\right)
+1
\right)\,. 
\end{equation}
Taking into account $N_\textrm{S}\gg1$ and $\frac{1}{\tau_\textrm{p}}\gg\frac{1}{\tau_\textrm{R}}$:
\begin{equation}
\kappa_\textrm{in} P_\textrm{p} = \frac{1}{\tau_\textrm{S}\tau_\textrm{p}\Gamma_\textrm{R}^\textrm{cav}}\,. 
\end{equation}
With $\kappa_\textrm{in}=\eta/(\hbar\omega_\textrm{p})$:
\begin{equation}
\frac{\eta}{\hbar\omega_\textrm{p}} P_\textrm{p} = \frac{\omega_\textrm{S}\omega_\textrm{p}}{Q_\textrm{S}Q_\textrm{p}}
\frac{n_\textrm{S}n_\textrm{p}V_\textrm{R}}{2\hbar\omega_\textrm{p}c^2g_\textrm{R}^\textrm{B}}
\frac{Q_\textrm{S}+Q_\textrm{R}}{Q_\textrm{S}}\,. 
\end{equation}
We obtain the result for the lasing threshold:
\begin{equation}
P_\textrm{p} = \frac{1}{\eta}
\frac{2n_\textrm{S}n_\textrm{p}\pi^2}{\lambda_\textrm{S}^\textrm{cav}\lambda_\textrm{p}^\textrm{cav}g_\textrm{R}^\textrm{B}}
\frac{V_\textrm{R}(Q_\textrm{S}+Q_\textrm{R})}{Q_\textrm{S}^2Q_\textrm{p}}\,. 
\end{equation}
 Using the experimental values summarized in Table\,\ref{Tab:exp} yield $P_{\textrm{th}}=187.5\,\textrm{mW}$ as stated in the main text.
\begin{table}[htbp]
\centering
\caption{\bf Summary of experimental parameters}
\begin{tabular}{ccccc}
\hline
$\lambda_\textrm{p}^\textrm{cav}$ & $634.57\,\textrm{nm}$ & \hspace{0.5cm} & $\lambda_\textrm{S}^\textrm{cav}$ & $693.13\,\textrm{nm}$\\
$n_{\textrm{p}}$ & $2.4$ & & $n_{\textrm{S}}$ & $2.4$\\
$Q_{\textrm{p}}$ & $297\,000\pm500$ & & $Q_{\textrm{S}}$ & $6\,650\pm50$ \\
$Q_{\textrm{R}}$ & $8\,960\pm290$ & & $g_{\textrm{R}}^{\textrm{B}}$ & $\sim40\,\textrm{cm/GW}$\,\cite{Savitski2013}\\
$\eta$ & $0.45$ & & $V_{\textrm{R}}$ & $109.85\,\upmu\textrm{m}^3$\\
\hline
\end{tabular}
 \label{Tab:exp}
\end{table}

\iffalse
%%%%%%%%%%%%%%%%%%%%%%%%%%%%%%%%%%%%%%%%%%%%%%%%%%%%%%%%%%%%%%%%%%%%%%%%%%%
% FIGURE XX - Raman Flower and Threshold
%%%%%%%%%%%%%%%%%%%%%%%%%%%%%%%%%%%%%%%%%%%%%%%%%%%%%%%%%%%%%%%%%%%%%%%%%%%
\begin{figure}[!tb]
\includegraphics[width=\columnwidth]{Figures/Fig_8_Q_Raman_Flower.pdf}
\caption{\textbf{(a)} Polarized Raman spectra obtained by rotating a $\frac{\lambda}{2}$-plate in front of the objective lens. From the angular dependence, we extract $\frac{I_{\textrm{min}}}{I_{\textrm{max}}}=0.61$.
\textbf{(d)} Calculation of the lasing threshold as a function of the bulk Raman gain coefficient $g_{\textrm{R}}^{\textrm{B}}$, calculated according to Eq.\,\ref{eq:threshold}.}
\label{fig:Flower}
\end{figure}
%%%%%%%%%%%%%%%%%%%%%%%%%%%%%%%%%%%%%%%%%%%%%%%%%%%%%%%%%%%%%%%%%%%%%%%%%%%
\fi

%%%%%%%%%%%%%%%%%%%%%%%%%%%%%%%%%%%%%%%%%%%%%%%%%%%%%%%%%%%%%%%%%%%%%
% FUTURE DIRECTIONS
%%%%%%%%%%%%%%%%%%%%%%%%%%%%%%%%%%%%%%%%%%%%%%%%%%%%%%%%%%%%%%%%%%%%%%%%%%%
\subsection{Future directions}
\label{sec:outlook}
We now turn to discuss some limiting factors and further possible improvements to this experiment. The double resonance condition is satisfied for the combination of $t_{\textrm{a}}$ and $t_{\textrm{d}}$ for which both pump and Stokes modes are resonant simultaneously. With the current top mirror design (depth of crater, $d=1.65\,\upmu\textrm{m}$) and diamond thickness $t_{\textrm{d}}=754.58\,\textrm{nm}$, a relatively large air-gap of $t_{\textrm{a}}=4173.98\,\textrm{nm}$ is required to meet this condition for the range of $\lambda_{\textrm{pump}}$ available. The large air-gap results in a large $V_{\textrm{R}}$, and consequently a large lasing threshold. Establishing the double resonance condition for a shorter air-gap will reduce $V_{\textrm{R}}$ and consequently $P_{\textrm{th}}$. 

\begin{figure}[!tb]
\includegraphics[width=\columnwidth]{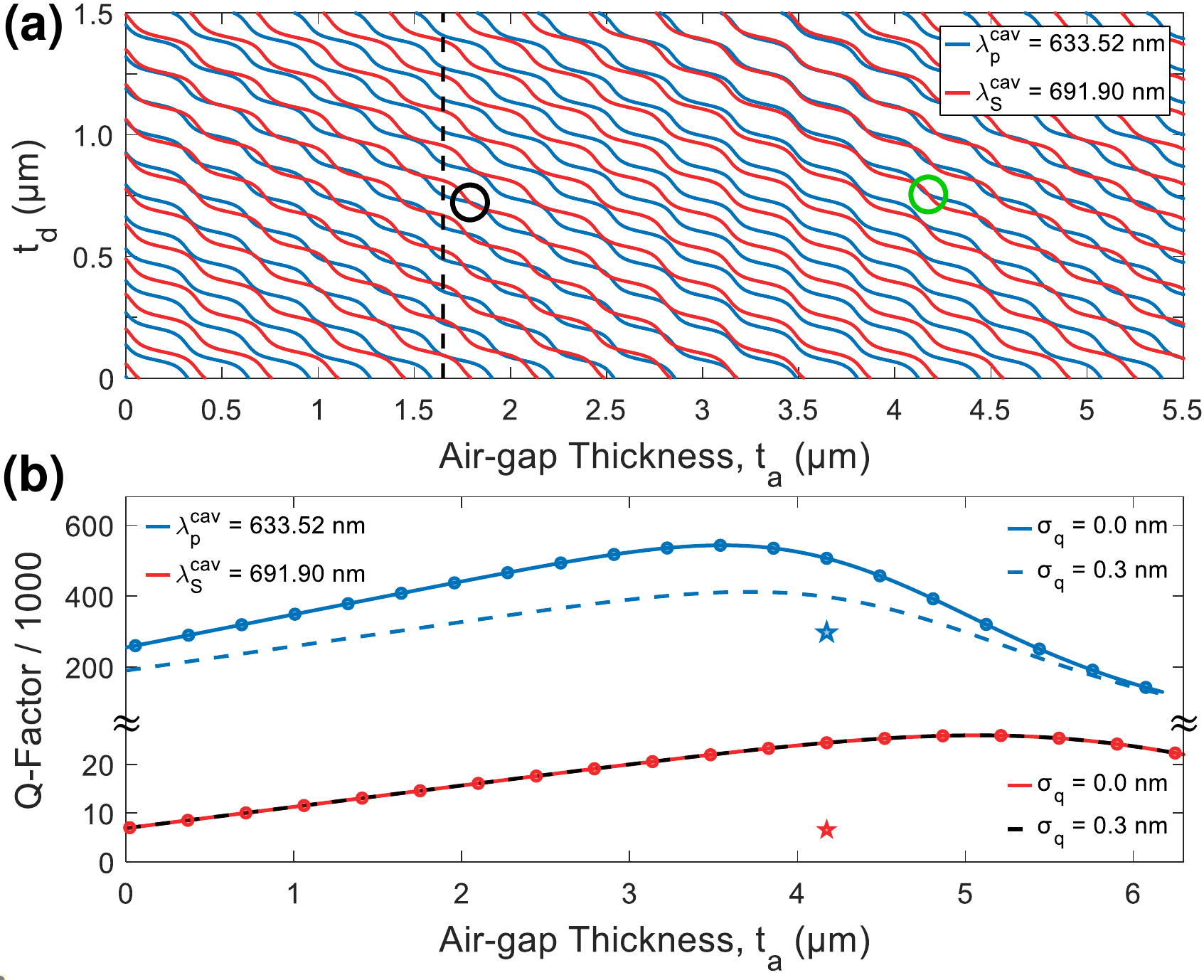}
\caption{\textbf{(a)} Simulated cavity mode structure for $\lambda_\textrm{p}^\textrm{cav}=633.52\,\textrm{nm}$ (blue) and $\lambda_\textrm{S}^\textrm{cav}=691.90\,\textrm{nm}$ (red) as a function of $t_{\textrm{a}}$ and $t_{\textrm{a}}$.
The double resonance condition is satisfied for the pair of $t_{\textrm{a,d}}$ where the two respective modes cross. The green circle highlights the combination $t_{\textrm{a}}=4173.98\,\textrm{nm}$ and $t_{\textrm{d}}=754.58\,\textrm{nm}$ used in this experiment. The black dashed line represents the depth of the crater. This depth sets a lower limit on the possible $t_{\textrm{a}}$ in the current cavity geometry. 
\textbf{(b)} Simulated dependence of the $Q$-factor with air-gap thickness for fixed $\lambda_\textrm{p}^\textrm{cav}=633.52\,\textrm{nm}$ (blue) and $\lambda_\textrm{S}^\textrm{cav}=691.90\,\textrm{nm}$ (red). The solid lines represent the $Q$-factor in the absence of any losses, while the dashed lines represent the $Q$-factor in the presence of surface scattering with surface roughness $\sigma_{\textrm{q}}=0.3\,\textrm{nm}$. The drop in $Q$-factors at large air-gap thicknesses is attributed to clipping losses at the top mirror. The experimentally measured $Q$-factors are indicated by the stars.} 
\label{fig:Spag_Q}
\end{figure}

We simulate the cavity for a wide range of $t_{\textrm{a}}$ and $t_{\textrm{d}}$ using $\lambda_\textrm{p}^\textrm{cav}=633.52\,\textrm{nm}$ and $\lambda_\textrm{S}^\textrm{cav}=691.10\,\textrm{nm}$ as before (Fig.\,\ref{fig:Spag_Q}\,(a)). Reducing the diamond thickness to $t_{\textrm{d}}=721.82\,\textrm{nm}$ satisfies the double-resonance condition for $t_{\textrm{a}}=1789.75\,\textrm{nm}$. For this air-gap, we calculate $V_{\textrm{R}}=20.42\,\upmu\textrm{m}^3$.

%%%%%%%%%%%%%%%%%%%%%%%%%%%%%%%%%%%%%%%%%%%%%%%%%%%%%%%%%%%%%%%%%%%%%%%%%%%
% FIGURE 9 - spaghetti and Q
%%%%%%%%%%%%%%%%%%%%%%%%%%%%%%%%%%%%%%%%%%%%%%%%%%%%%%%%%%%%%%%%%%%%%%%%%%%
%%%%%%%%%%%%%%%%%%%%%%%%%%%%%%%%%%%%%%%%%%%%%%%%%%%%%%%%%%%%%%%%%%%%%%%%%%%

An additional benefit of reducing $t_{\textrm{a}}$ becomes apparent on simulating the behavior of the $Q$-factor with increased cavity length (Fig.\,\ref{fig:Spag_Q}\,(b)). The $Q$-factor and the cavity round-trip loss $\mathcal{L}_{\textrm{cav}}$ are linked via $Q=\frac{4\pi L_{\textrm{cav}}}{\lambda\mathcal{L}_{\textrm{cav}}}$, where $L_{\textrm{cav}}=t_{\textrm{a}}+L_{\textrm{0}}$. Here the term $L_{\textrm{0}}$ describes the diamond thickness and the field penetration into the DBR mirror coatings\,\cite{Koks2021,Flagan2021}. For short cavity lengths, the $Q$-factor increases linearly with $t_{\textrm{a}}$. However, for large cavity lengths, the extent of the intensity mode waist at the top mirror, $w_{\textrm{I}}$, becomes larger than the spherical extent of the mirror, leading to beam clipping and a subsequent drop in the $Q$-factor\,\cite{Hunger2010,Flagan2021,Benedikter2015}. For a spherical mirror with diameter $D$, the clipping losses are calculated according to $\mathcal{L}_{\textrm{clip}}=e^{-\nicefrac{D^2}{2w_{\textrm{I}}^2}}$, where the beam waist $w_{\textrm{I}}$ evolves according to\,\cite{Flagan2021}
\begin{equation}
w_{\textrm{I}}=\sqrt{\frac{\lambda R_{\textrm{cav}}}{\pi}}\times\left(\frac{R_{\textrm{cav}}}{(t_{\textrm{a}}+\frac{t_{\textrm{d}}}{n_{\textrm{d}}})}-1\right)^{-\frac{1}{4}}\,. 
\end{equation}
In Fig.\,\ref{fig:Spag_Q}\,(b), a drop in $Q$-factor is expected for ${t_{\textrm{a}}\gtrsim3.5\,\upmu\textrm{m}}$, a consequence of clipping losses. Therefore, a shorter $t_{\textrm{a}}$ will have the added benefit of preserving a high $Q$-factor. Here, the value of $R_{\textrm{cav}}=11\,\upmu\textrm{m}$ and $D=6\,\upmu\textrm{m}$ are extracted from a scanning confocal microscope (Keyence Corporation). 

Using this model for $t_{\textrm{d}}=721.82\,\textrm{nm}$ and $t_{\textrm{a}}=1789.75\,\textrm{nm}$ (black circle in Fig.\,\ref{fig:Spag_Q}\,(a)), we find a theoretical $Q_{\textrm{p}}=406\,060$, $Q_{\textrm{S}}=13\,550$, $\eta=0.81$ and consequently $P_{\textrm{th}}=4.94\,\textrm{mW}$.

The diamond surface introduces scattering losses which should be taken into account. Surface scattering can be incorporated in the transfer-matrix simulations according to Ref.\,\cite{Carniglia2002}. Motivated by typical roughness measurements reported by Ref.\,\cite{Appel2016,Flagan2021}, including a scattering layer with surface roughness ${\sigma_{\textrm{q}}=0.3\,\textrm{nm}}$, reduces the $Q$-factor to $Q_{\textrm{p}}^{\textrm{scat}}=258\,780$ and $Q_{\textrm{S}}^{\textrm{scat}}=13\,480$. Consequently, the additional loss-channel reduces $\eta^{\textrm{scat}}=0.59$.
The reduction in the $Q$-factor increases the lasing threshold to $P_{\textrm{th}}^{\textrm{scat}}=10.71\,\textrm{mW}$. Finally, increasing the thickness of the diamond membrane provides a way to reduce further the lasing threshold on the account of a larger $Q$-factors offered by the longer effective cavity length. Applying the same method as in Fig.\,\ref{fig:Spag_Q}\,(a), we find that the double-resonance condition is established for $t_{\textrm{d}}=3361\,\textrm{nm}$ and $t_{\textrm{a}}=1726\,\textrm{nm}$. Using these values and simulating a loss-less cavity yields $P_{\textrm{th}}=0.78\,\textrm{mW}$. Including surface scattering ($\sigma_{\textrm{q}}=0.3\,\textrm{nm}$) increases the threshold to $P_{\textrm{th}}=1.87\,\textrm{mW}$.

\end{document}